%% file: bartinfluence.tex
\documentclass[12pt]{article}
\usepackage{graphicx,color}
\usepackage{amsmath}
\usepackage{amsfonts}
\usepackage{natbib}
\usepackage{geometry}
\usepackage{units}
\usepackage{caption}
\usepackage{url}
\usepackage{dutchcal} % for lowercase script-m, which is used for minnumbot.
\usepackage{mathabx} % for \Asterix
\newcommand{\blind}{0}

\def\C {\,|\:}
\def\x {\bf x}
\def\I {\mathrm{I}}
\def\minbot {n_0}

\begin{document}

\if0\blind
{
	\title{Influential Observations in Bayesian Regression Tree Models}	
	\author{
	M.~T.~Pratola\thanks{Department of Statistics, The Ohio State University, 1958 Neil Avenue, 404 Cockins Hall, Columbus, OH 43210-1247 (mpratola@stat.osu.edu).},
	E.~I.~George\thanks{Department of Statistics, The Wharton School,  University of Pennsylvania.},
	and
	R.~E.~McCulloch\thanks{School of Mathematical and Statistical Sciences, Arizona State University.}
	}
%\date{\today}
} \fi

\if1\blind
{
  \bigskip
  \bigskip
  \bigskip
	\title{Influential Observations in Bayesian Regression Tree Models}	
  \medskip
} \fi

\maketitle
\thispagestyle{empty}

\begin{abstract}
\input{abs.tex}

\end{abstract}

\newpage

\setcounter{page}{1}
%\tableofcontents
%\thispagestyle{empty}

%\newpage

% Introduction
\section{Introduction}
\label{section:intro}
\input{intro.tex}

\section{Bayesian Regression Trees and BART}
\label{section:background}
\input{model.tex}

% Diagnostics
\section{Influence Diagnostics for Trees}
\label{section:diagnostics}
\input{diagnostics.tex}

% Importance Sampling for Robust Prediction
\section{Adjusting Predictions via Importance Sampling}
\label{section:importance}

\input{importance.tex}

% Robustifying BART
%\section{Robust BART}
%\input{robustbart.tex}

% Examples
\section{Examples}
\label{section:examples}

\input{examples.tex}

% Discussion
\section{Conclusion}
\label{section:conc}
\input{conclusion.tex}

\if0\blind
{
\section*{Acknowledgements}
The work of MTP was supported in part by the National Science Foundation (NSF) under Agreement DMS-1916231 and in part by the King Abdullah University of Science and Technology (KAUST) Office of Sponsored Research (OSR) under Award No. OSR-2018-CRG7-3800.3.  The work of EIG was supported by NSF DMS-1916245.  The work of REM was supported by NSF DMS-1916233.
}\fi

%\section*{Supplementary Materials}
%The following supplementary materials are available online in the single archive file supplement.zip.
%\begin{description}
%	\item[Supplement to Heteroscedastic BART Via Multiplicative Regression Trees:] Additional plots and examples of using HBART. (supplement.pdf)
%	\item[code.zip:] R script demonstrating the application of R-package ``rbart'' on the simulated data example. (code.zip, GNU zipped file)
%\end{description}

\clearpage

\bibliography{./bartinfluence.bib}
\bibliographystyle{./jasa}
%\bibliographystyle{./jasa}

% Online Supplement -- proofs.
\if0
{
\newpage
\setcounter{page}{1}
\section*{Supplement}\label{section:supp}
\input{supp.tex}
}\fi

\end{document}

%% file: abs.tex
{ BCART (Bayesian Classification and Regression Trees) and BART (Bayesian Additive Regression Trees) are popular Bayesian regression models widely applicable in modern regression problems.  Their popularity is intimately tied to the ability to flexibly model complex responses depending on high-dimensional inputs while simultaneously being able to quantify uncertainties.  This ability to quantify uncertainties is key, as it allows researchers to perform appropriate inferential analyses in settings that have generally been too difficult to handle using the Bayesian approach.  However, surprisingly little work has been done to evaluate the sensitivity of these modern regression models to violations of modeling assumptions.  In particular, we will consider influential observations, which one reasonably would imagine to be common -- or at least a concern -- in the big-data setting.  In this paper, we consider both the problem of detecting influential observations and adjusting predictions to not be unduly affected by such potentially problematic data.  We consider three detection diagnostics for Bayesian tree models, one an analogue of Cook's distance and the others taking the form of a divergence measure and a conditional predictive density metric, and then propose an importance sampling algorithm to re-weight previously sampled posterior draws so as to remove the effects of influential data in a computationally efficient manner.  Finally, our methods are demonstrated on real-world data where blind application of the models can lead to poor predictions and inference.
}
\\

%% file: intro.tex
In the contemporary approach to data-driven problem solving, statistical models have received increasing attention and popularity as a means for arriving at answers to complex research, science and business questions. As datasets have increased in size with the transition to the  ``big-data'' era, the complexity and scalability of statistical models have seen rapid advances in order to address the needs of these modern problems.  Popular examples of such models include neural networks \citep{ghugare:etal:2014}, random forests \citep{leo:2001} and localized Gaussian Processes \citep{gramacy:2015}.  In problems where uncertainty quantification is deemed necessary, Bayesian methods have come to the fore, such as the Bayesian variants of neural networks \citep{mackay1995probable}, Bayesian localized GPs \citep{liu:etal:2021} and Bayesian Regression Tree models \citep{chipman:etal:1998,chipman:etal:2010,Pratola:2016,horiguchi:etal:2021,horiguchi:etal:2022}.

Despite the increasing popularity and capability of these modern statistical tools, there has been a conspicuous disconnect in terms of tools that support the application of such complex models when compared to their humble, small-dataset, low-dimensional ancestors.  For example, in linear regression, students are taught an extensive array of tools for validating modeling assumptions in the classical setting, such as residual diagnostics, outlier detection and influence metrics \citep{weisberg2013applied,cook:weisberg:1982}.   The Bayesian linear model has also received attention earlier in the  literature \citep{chaloner:brant:1988,zellner:moulton:1985,johnson:geisser:1983,zellner:1975}. Yet surprisingly, such supporting tools have not received the same attention in the development of modern variants of statistical models.  The assumption, it seems, is that in the big-data setting such issues are of lesser concern.  We have found this assumption to be incorrect.

Our focus in this paper is on the single-tree Bayesian classification and regression tree (BCART) model \citep{chipman:etal:1998}, and the ensemble-of-trees Bayesian Additive Regression Tree (BART) model of \cite{chipman:etal:2010} in particular.  This class of models is currently receiving much attention in the research community, and has been used in a wide variety of problems including medical studies \citep{tan2019bayesian}, causal analysis \citep{hahn2020bayesian}, computer experiments \citep{Pratola:etal:2014} and applied optimization \citep{horiguchi:etal:2022}.  BCART and BART models have contributed to this popularity due to their ability to scale to moderately sized big-data applications while retaining the ability to fully quantify statistical uncertainties due to the elegant exploitation of conjugacies in the MCMC sampler. % as well as developments to efficiently sample the so-called tree-space \citep{Pratola:2016}.  
Our work arose out of a simple curiosity: can BCART or BART models be negatively affected by a potentially problematic observation, i.e. an observation that can be influential or is an outlier (or both)?  On the one hand, since such Bayesian tree models  fit simple localized models, it may  appear that any problematic behavior due to a bad observation would be localized, and perhaps of not serious concern when working with large datasets.  On the other hand, big-data usually is also high-dimensional, and in high-dimensions our notions of what constitutes a large sample size may not match our intuition.

\begin{figure}[ht]
\centering
\includegraphics[scale=.28]{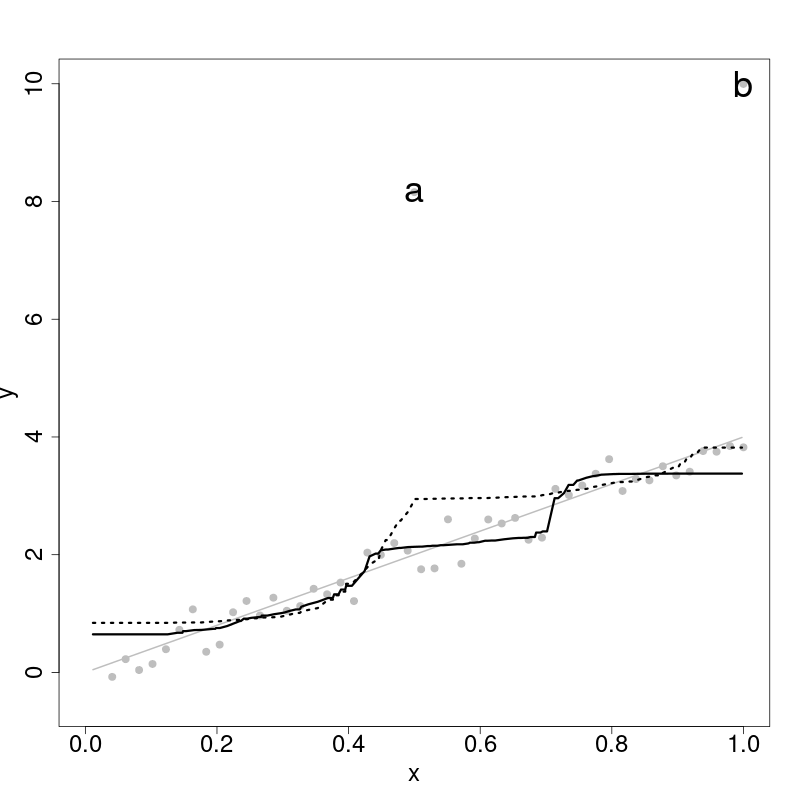} %cookd-sim-1x-refit-t1.pdf}
\includegraphics[scale=.28]{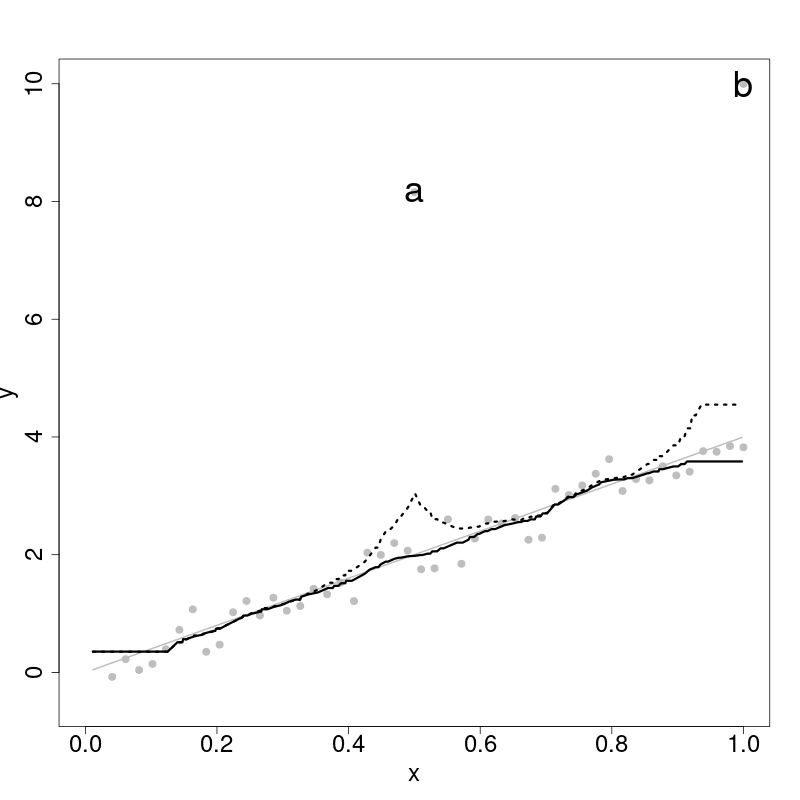} %cookd-sim-1x-refit-tm.pdf}
\caption{Effect of two problematic observations on posterior predictions of BCART (left panel, $m=1$ trees) and BART (right panel, $m=200$ trees) models.  Observation `a' is a large outlier but has less influence due to its location in the center of the regression domain, while observation `b' is both a large outlier and has higher influence due to its location at the edge of the regression domain.  The resulting fits demonstrate the effect of removing these observations on the resulting posterior predictions (solid line) versus leaving them in (dotted line). The grey solid line denotes the true mean function.}
\label{fig:refit1d}
\end{figure}

Figure \ref{fig:refit1d} demonstrates a simple example of this scenario.  In fact, this scenario is rather favorable as it is low-dimensional and there is plenty of data -- both of these choices make it easy to visualize the behavior of BART with $m=1$ and $m=200$ trees.  Yet despite this seemingly favorable situation (i.e. one where we might {\em not} expect any serious effect due to influence or outliers), there is a suprisingly strong effect of two problematic observations, denoted in the figure by the `a' and `b' symbols.  Certainly, the effect is localized as expected, so the overall fit may be reasonable for much of the regression domain of interest.  Yet, in the local region containing the problematic observation, the predictions are severely affected and it is reasonable to expect this type of issue to become worse in more realisitic, higher-dimensional applications.

In the classical linear regression model, $y_i=x_i\beta+\epsilon_i, \ \ \epsilon_i\sim N(0,\sigma^2)$ where $x_i\in\mathbb{R}^d,$ the approach to handling such problematic observations is to detect and remove such observations before proceeding to the final model fitting and inference stages.  The popular classical tools include calculating the leverage of observations based on the diagonal entries of the hat-matrix, and calculating Cook's distance, defined for observation $i$ as
\[ 
D_i=\frac{\sum_{j=1}^n\left(\hat{y}_j-\hat{y}_{j(i)}\right)^2}{ds^2}
\]
where $\hat{y}_{j(i)}$ represents the model's prediction when observation $i$ is held out from the training data, and $s^2$ is the usual least-squares estimate of error variance, $\sigma^2$.
Cook's distance can itself can be factored into a term that represents detecting observations problematic due to a large potential for influence (leverage) and a term detecting influence due to a large residual.  These terms are combined product-wise to arrive at $D_i,$ implying that good observations are those with low residual and low leverage while problematic observations could be problematic for either or both of these issues.

In the modern Bayesian context, it is less clear how to handle such problematic data.  For instance, do we care about point predictions or do we care about the posterior distribution?  In the former setting, an analogue of the classical Cook's distance may be quite reasonable.  In the latter setting, some theoretical work suggests that if the problematic observation is known, one may not need to completely remove it from the analysis; instead the posterior can be adjusted to correct the effect of the problematic data on the resulting posterior.  This implies that there are in fact two procedures needed to appropriately handle problematic observations in our modern Bayesian regression setting:
\begin{enumerate}
\item[i.] identification of problematic observations;
\item[ii.] model (posterior) adjustment given identified problematic observations.
\end{enumerate}
In this work, we propose three approaches for the identification problem (i).  First, a direct extension of Cook's distance to the regression tree model setting is outlined, and has the benefit of providing an easy and sensible interpretation.  Second, an alternative divergence-based metric is also proposed.  The divergence approach has the benefit of identifying observations that affect the posterior distribution.  Third, identification can be performed by detecting changes in the conditional predicitve distribution.  For the adjustment problem (ii), we explore two alternatives: the simple (but wasteful) dropped-observation approach, and an importance-sampling approach that reweights posterior expectations to account for the problematic observation without going so far as to completely remove it.

The paper proceeds as follows.  
In Section \ref{section:background} we review the the BCART and BART models. 
In Section \ref{section:diagnostics}, we outline our proposed Cook's distance metric for trees as well as the divergence and predictive distribution metrics.  In Section \ref{section:importance} we derive importance samplers for reweighting BCART and BART posteriors to account for problematic observations.  %In Section \ref{section:robust} we use the reweighting schemes to robustify BART predictions (maybe).  
We then apply these tools to a variety of simulated datasets and to real-world data involving biomass fuels in Section \ref{section:examples}. Finally, we conclude with a discussion in Section \ref{section:conc}.

%% file: model.tex
For high-dimensional regression, most statistical  and machine learning techniques focus on the estimation of $E[y\C\x]=f(\x)$.
%When a model for conditional variance is explicitly considered, 
It is typically assumed that $Var[y \C \x]=\sigma^2$ with
%A commonly assumed scenario for many statistical regression and machine learning techniques is that of modeling a process $Y({\bf x})$ that is formed by an unknown mean function, $E[Y\vert {\bf x}]=f({\bf x}),$ and an unknown constant variance, $Var[Y\vert {\bf x}]=\sigma^2,$ along with a stochastic component arising from independent random perturbations $Z.$  
the data generated according to the homoscedastic process
\begin{align}
\label{eq:homoprocess}
y({\bf x}) = f({\bf x})+\sigma Z
\end{align}
where $Z\sim\text{N}(0,1)$ and ${\bf x}=(x_1,\ldots,x_d)$ is a $d$-dimensional vector of predictor variables.  %Such a process is known as a homoscedastic process.  

%When combined with a prior, each such regression tree provides a simple binary recursive partition of the multidimensional predictor space, which is adaptively learned from the data. 
%provide a simple yet powerful of non-parametric specifying adaptive regression bases where, the basis elements themselves, when combined with a prior, adapt to the observed data.}
 
\begin{figure}[ht!]
\begin{center}
\includegraphics[scale=.5]{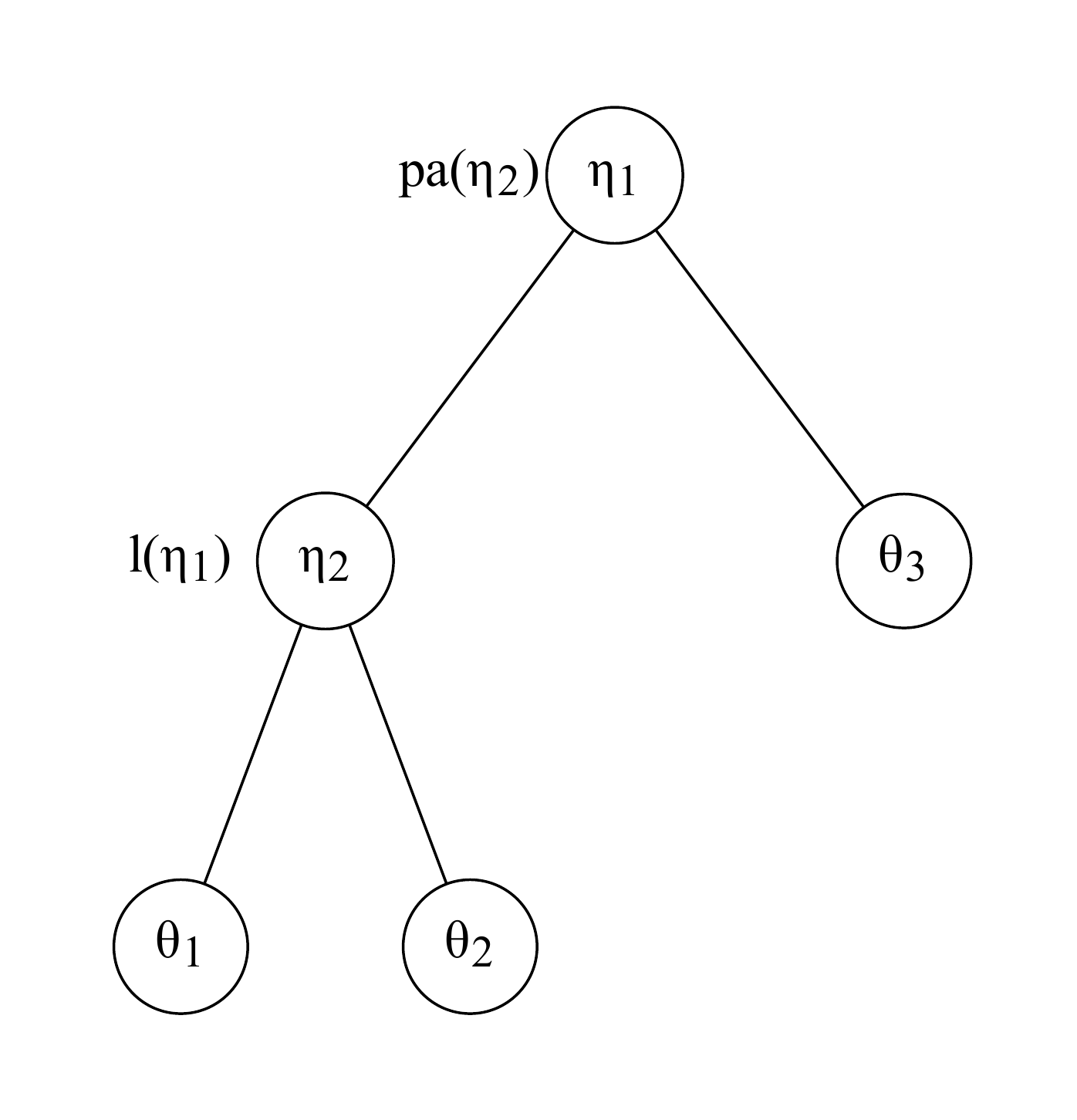}
\end{center}
\caption{Labeling for a single regression tree ${\bf T}$.  Nodes are denoted by circles and labeled using the symbol $\eta$.  Lines denote branches connecting the nodes.  Nodes can also be identified as left and right children (e.g. $\eta_2=l(\eta_1)$) or as parent (e.g. $\eta_1=pa(\eta_2)$).  Terminal nodes have no branches below them and contain associated parameter values $\theta$.  %A sub-tree uses the $T$ symbol with subscript given by the node index (e.g. the subtree including $\eta_3$ and all its children is $T_3$).  
Note that later in the paper ${\bf T}_j$ will also index one member of an ensemble of trees, its use will be clear from context.}
\label{fig:treelabels}
\end{figure}

BART models the unknown mean function $f(\x)$ with an ensemble of Bayesian regression trees.  Such regression trees provide a simple yet powerful non-parametric specification of multidimensional regression bases, where the form of the basis elements are themselves learned from the observed data.
Each Bayesian regression tree is a recursive binary tree partition that is made up of interior nodes, ${\bf T}$, and a set of parameter values, ${\bf M}$, associated with the terminal nodes.  Each interior tree node, $\eta_i$, has a left and right child, denoted $l(\eta_i)$ and $r(\eta_i)$.  In addition, all nodes also have one parent node, $pa(\eta_i),$ except for the tree root.  One may also refer to a node by a unique integer identifier $i$, counting from the root where the left child node is $\eta_{2i}$ and the right child node is $\eta_{2i+1}$.  For example, the root node $\eta_1$ is node $1$, with node $2$ and node $3$ being $\eta_1$'s children.  %One can also label a subtree starting at node $\eta_i$ simply as $T_i$.  
Figure \ref{fig:treelabels} summarizes our notation.

Internal nodes of regression trees have split rules depending on the predictors and ``cutpoints'' that are the particular predictor values at which the internal nodes split.  This modeling structure is encoded in ${\bf T}$, which accounts for the split rules at each internal node of a tree and the topological arrangement of nodes and edges forming the tree.  Given the design matrix ${\bf X}$ of predictors having dimension $n\times d$, each column represents a predictor variable $v \in \lbrace 1,\ldots,d\rbrace,$ and each row ${\bf x}$ corresponds to the observed settings of these predictors. At a given internal node, the split rule is then of the form $x_v<c$ where $x_v$ is the chosen split variable and $c$ is the chosen cutpoint $c$ for split variable $x_v.$

The Bayesian formulation proceeds by specifying discrete probability distributions on the split variables $v$ taking on a value in $\{1,\ldots,d\}$ and specifying discrete probability distributions on the set of distinct possible cutpoint values, where $n_v$  is the total number of discrete cutpoints available for variable $v$.  For a discrete predictor, $n_v$ will equal the number of levels the predictor has, while for a continuous predictor a choice of $n_v=100$ is common \citep{chipman:etal:2010}.  The internal modeling structure of a tree, ${\bf T},$ can then be expressed as ${\bf T}=\{(v_1,c_1),(v_2,c_2),\ldots\}$.

The Bayesian formulation is completed by specifying prior distributions on the parameters at the terminal nodes. For $B=|{\bf M}|$ terminal nodes in a given tree, where $|\cdot|$ denotes cardinality, the corresponding parameters are ${\bf M}=\{\mu_1,\ldots,\mu_{B}\}$. Taken all together, the Bayesian regression tree defines a function $g({\bf x};{\bf T},{\bf M})$ which maps input ${\bf x}$ to a particular $\mu_j,$ $j\in 1\ldots B.$

%In the BART methodology, the model specifies the use of an ensemble of regression trees by combining a sum of the basic binary regression tree arrangement outlined above under the constant variance assumption,

The original BART model is then obtained as the ensemble sum of $m$ such Bayesian regression trees plus an error component, 
$%\[
y({\x}_i) = \sum_{j=1}^m g({\x}_i;{\bf T}_j,{\bf M}_j) + \sigma Z_i, \ \ Z_i \sim N(0,1)
,\ \ $%\]
where $y({\x}_i)$ is the observation collected at predictor setting $\x_i$, and $\sigma^2$ is the variance of the homoscedatic process.  Combining all the parameters together as $\boldmath{\Theta}=({\bf T},{\bf M},\sigma^2)$, the BART prior is factored as $\pi(\boldmath{\Theta})=\pi({\bf M}\vert{\bf T})\pi({\bf T})\pi(\sigma^2).$  For the terminal node parameters, normal priors are specified as,
$%\[
\pi(\mu_{jk}) \sim N(0,\tau^2),
\ $%\]
where $\mu_{jk}$ is the $k$th terminal node component for tree $j$, and an inverse chi-squared prior is specified for the variance,
$%\[
%\sigma^2 \sim \frac{\nu\lambda}{\chi^2_\nu} \equiv \chi^{-2}(\nu,\lambda).
\sigma^2 \sim \chi^{-2}(\nu,\lambda),
\ $%\]
where 
$\chi^{-2}(\nu,\lambda)$ denotes the distribution
%$\frac{\nu \lambda}{\chi^2_\nu}$.
$(\nu \lambda)/\chi^2_\nu,$ and $\chi^2_\nu$ is the chi-squared distribution with $\nu$ degrees of freedom.
For a prior on the tree structure, we specify a stochastic process that describes how a tree is drawn.
A node at depth $\delta\in\lbrace 0,1,2,\ldots\rbrace$ spawns children with probability
$%\[
\alpha(1+\delta)^{-\beta},
\ $%\]
for $\alpha\in (0,1)$ and $\beta\geq 1.$
As the tree grows, $\delta$ gets bigger so that a node is less likely to spawn children and
more likely to remain a terminal node, thus penalizing tree complexity.
Details on specifying the parameters of the prior distributions are discussed 
in detail in \cite{chipman:etal:2010}, while typically the choice $m=200$ 
trees appears to be reasonable in many situations \citep{chipman:etal:2010,hill2011bayesian,starling2020bart,horiguchi:etal:2021}.  Meanwhile, choosing $m=1$ results in the BCART model.

The use of normal priors on the terminal node $\mu$'s, and an inverse chi-square prior on the  variance, greatly facilitates the posterior simulation via an MCMC algorithm as they are conditionally conjugate.
Selecting the split variables and cutpoints of internal tree nodes is performed using a Metropolis-Hastings step by growing and pruning each regression tree. The growing/pruning are performed using so-called birth and death proposals, which either split a current terminal node in ${\bf M}$ on some variable $v$ at some cutpoint $c$, or collapse two terminal nodes in ${\bf M}$ to remove a split.  For complete details of the MCMC algorithm, the reader is referred to \cite{chipman:etal:1998, denison:etal:1998,chipman:etal:2010,Pratola:2016}.

%% file: diagnostics.tex
We now outline three diagnostic tests for the detection of problematic observations.  The first is a direct application of Cook's distance to Bayesian regression trees,  the second is a divergence-based approach and the third a conditional predictive distribution approach.

\subsection{Conditional Cook's Distance for Regression Trees}

Conditioning on the tree $(T,M)$, a single regression tree can be expressed in the usual linear form as $g(x;T,M)=\sum_{b=1}^B\mu_b\I_b(x)$ where $B$ is the total number of terminal nodes in the tree and $\I_b(x)$ is the indicator function taking the value $1$ when $x$ maps to the hyperrectangle defined by terminal node $b$, and $0$ otherwise.  The analogous formula for Cook's distance in the single-tree case by regressing $y$ on $I_b(x)$ (see Supplement) becomes
\begin{align}
\label{eqn:singletreecooks}
D_i &= \underbrace{\frac{1}{B}}_\text{Tree Complexity}\times\underbrace{\left(\frac{e_i}{\sigma}\right)^2}_\text{Normalized Residual}\times\underbrace{\frac{n_{(i)}}{(n_{(i)}-1)^2}}_\text{Node Purity}
\end{align}
where $e_i=y_i-\sum_{b=1}^B\mu_bI_b(x_i)$ is the regression residual for observation $i,$ and $n_{(i)}$ is the number of observations in the terminal node to which observation $i$ maps.  Note here that in comparison to the classical Cook's distance, we have replaced $\hat{\sigma}$ with the parameter itself, for which we have samples.  This form of $D_i$ provides helpful interpretations.  For instance, it is a decreasing function of the number of terminal nodes, $B,$ but on the other hand it increases as node purity increases (i.e. as $n_{(i)}$ becomes small) and in particular will blow up when $n_{(i)}-1=0.$  %This behavior reflects the fact that a model with arbitrarily large $B$ is hardly a good model.  
Also, we see that $D_i$ increases if the residual of observation $i$ is large relative to the standard error, and this effect increases like the square for every unit increase in standard devation of the residual for observation $i$.  To arrive at an overall estimate, we take the posterior sample mean over our $N$ MCMC draws of $(T,M,\sigma)$,
\begin{align}
\label{eqn:postsingletreecooks}
\widehat{E[D_i\vert {\bf Y}]} &= \frac{1}{N}\sum_{k=1}^ND_i^{(k)}
\end{align}
where each $D_i^{(k)}$ is the conditional Cook's distance as defined in equation (\ref{eqn:singletreecooks}).

For the sum-of-trees BART model, we can extend this idea in a few ways.  One simple approach is to report the average $D_i$ across all of the $m$ tree's in BART's sum. That is, if $D_{ji}^{(k)}$ is the Cook's distance calculated as in equation (\ref{eqn:singletreecooks}) above for tree $j$, then one could report 
$$\widehat{E[\overline{D}_i\vert{\bf Y}]} = \frac{1}{N}\sum_{k=1}^N\overline{D}_i^{(k)} \text{ where } \overline{D}_i^{(k)}=\frac{1}{m}\sum_{j=1}^mD_{ji}^{(k)}.$$ 
Another practical alternative would be to report the average maximum $D_{ji}$ over the trees, $$\widehat{E[D_i^\star \vert {\bf Y}]}=\frac{1}{N}\sum_{k=1}^ND_i^{\star(k)} \text{ where } D_i^{\star(k)}=max_jD_{ji}^{(k)}.$$  The exact solution can be found by converting each sum-of-trees function into a single tree representation as in \cite{horiguchi:etal:2021} to calculate the conditional Cook's distances for the BART sum-of-trees ensemble.  In this case, the linear form is expressed as $g(x;T^S,M^S)=\sum_{b=1}^{B^S}\mu_b\I_b(x)$ where the superscript $S$ denotes the supertree representation of the ensemble. Note that each $\mu_b$ here is itself the sum of $m$ $\mu_{jk}$ parameters from the original BART representation. If $B_j$ is the number of terminal nodes in tree $T_j,$ then the number of terminal nodes in the supertree $B^S<\sum_{j=1}^mB_j$, where typically this inequality is `$<<$'.  Let $n_{j,(i)}$ be the number of observations in the terminal node of tree $j$ to which observation $i$ maps, and similarly let $n^S_{(i)}$ be the number of observations in the supertree's terminal node to which observation $i$ maps.  Typically, $n^S_{(i)}<\min_j n_{j,(i)}<<\sum_{j=1}^m n_{j,(i)}$ since the hyperrectangle defined by the supertree terminal node to which observation $i$ maps is the intersection of the corresponding hyperrectangles from the $m$ trees in the additive form, i.e. $\text{vol}(\I_{b^S}(x_i)):=\text{vol}(\cap_{j=1}^m\I_{b_j}(x_i)).$ We can then calculate the conditional Cook's distance as $D_i=\frac{1}{B_S}\left(\frac{e_i}{\sigma}\right)^2\frac{n^S_{(i)}}{(n^S_{(i)}-1)^2}$  and then report the posterior average of these values.  Note that in this exact calculation, we see that it is likely for the influential or outlying observation $i$ to have a much smaller $n^S_{(i)}$ than any single tree $j$, which serves to inflate the Cook's distance more agressively than in the above approximations.  However, the approximations  $\overline{D}_i^{(k)}$ and $\widetilde{D}_i^{(k)}$ may be preferable for their computational simplicity.%, and in particular $\widetilde{D}_i^{(k)}$ since if either Cook's distance is dominated by the node purity term, then $\arg\max_jD_{ji}^{(k)}\equiv \arg\min_j n_{j,(i)}.$

\subsection{Kullback-Leibler Divergence Diagnostic}
Recall the Kullback-Leibler divergence from distribution $Q$ to $P$ is defined as
\[
D_{KL}(P\vert\vert Q):=\int_{-\infty}^{\infty}log\left(\frac{P}{Q}\right)dP
\]
where $D_{KL}\geq 0$ with equality iff $P=Q.$  In our context, we propose to take the reference distribution to be the posterior involving all the data, $$P:=\pi(\boldmath{\Theta}\vert {\bf Y}),$$
and the distribution $Q$ is taken to be the posterior when the potentially problematic data is held out.  If we consider the simplest case of holding out a single obseration $y_i$, then $$Q:=\pi(\boldmath{\Theta}\vert{\bf Y}_{-i}).$$  The KL divergence diagnostic has a simple Bayesian interpretation when evaluating the potential for observations to be problematic:  if $D_{KL}\approx 0$ then observation $y_i$ is not very influential on the posterior distribution, whereas if $D_{KL}>> 0$ then observation $y_i$ is unduly influential on the posterior distribution.

In practice, we can estimate this metric quite simply using posterior samples from our full-data fit.  Denoting $f(\cdot\vert\boldmath{\Theta})$ as the likelihood function, for the theoretical divergence we have
\begin{align}
\label{eqn:kldiv}
D_{KL}(\pi(\boldmath{\Theta}\vert {\bf Y})\vert\vert \pi(\boldmath{\Theta}\vert {\bf Y}_{-i})) &= \int_{\boldmath{\Theta}} \log\left(
\frac{f({\bf Y}\vert\boldmath{\Theta})\pi(\boldmath{\Theta})/\pi({\bf Y})}{f({\bf Y}_{-i}\vert\boldmath{\Theta})\pi(\boldmath{\Theta})/\pi({\bf Y}_{-i})}\right)\pi(\boldmath{\Theta}\vert{\bf Y})d\boldmath{\Theta} \nonumber \\
%&= \int_\Theta log\left(
%\frac{f({\bf Y}\vert\Theta)}{f({\bf Y}_{-i}\vert\Theta)}
%\right)\pi(\Theta\vert {\bf Y})d\Theta+\log\left(\frac{\pi({\bf Y}_{-i})}{\pi({\bf Y})}\right)\\
&= \int_{\boldmath{\Theta}}\log\left(f(y_i\vert\boldmath{\Theta})\right)\pi(\boldmath{\Theta}\vert{\bf Y})d\boldmath{\Theta} + \log\left(\frac{\pi({\bf Y}_{-i})}{\pi({\bf Y})}\right)
%&\approx \sum_{k=1}^N log\left(f(y_i\vert\theta^{(k)})\right)
\end{align}
where the first term is due to the i.i.d. form of the likelihood. Since $D_{KL}\geq 0$, we know that the divergence is minimized when both the first and second term are zero. The first term captures the contribution of $y_i$ to the posterior distribution of $\boldmath{\Theta}$, and can be 
approximated using the full-data posterior samples, $\boldmath{\Theta}^{(k)}\sim \pi(\boldmath{\Theta}\vert{\bf Y}),$ as $\frac{1}{N}\sum_{k=1}^N\log\left(f(y_i\vert\boldmath{\Theta}^{(k)})\right).$  The second term is a bit more involved, but can be easily estimated by recognizing it as $\log\left(\left[\pi(y_i\vert{\bf Y}_{-i})\right]^{-1}\right).$  Note the connection of this term to the Conditional Predictive Ordinate (CPO), defined as $\pi(y_i\vert {\bf Y}_{-i}),$ where large values of CPO indicate a good fitting model for $y_i$ while large values of the inverse of CPO identify problematic observations \citep{pettit1990conditional,gelfand1992model,gkisser2017predictive}. We can estimate this term appealing to an importance sampling trick as summarized in Proposition 0.

{\bf Proposition 0:} Suppose $\pi(\Theta\vert {\bf Y})$ and $\pi(\Theta\vert {\bf Y}_{-i})$ are probability density functions such that $\pi(\Theta\vert {\bf Y})>0$ whenever $\pi(\Theta\vert {\bf Y}_{-i})>0.$  Consider $\pi(y_i\vert {\bf Y}_{-i})=E_\Theta\left[\pi(y_i\vert{\bf Y}_{-i},\Theta)\pi(\Theta\vert{\bf Y}_{-i})/\pi(\Theta\vert{\bf Y})\right]$ where the expectation $E_\Theta$ is with respect to $\pi(\Theta\vert{\bf Y}).$  Let $\Theta^{(1)},\ldots,\Theta^{(N)}\sim \pi(\Theta\vert{\bf Y})$ be independent.  Then, $\log\left(\frac{1}{N}\sum_{k=1}^N\left[f(y_i\vert\Theta^{(k)})\right]^{-1}\right)\rightarrow \log\left(\frac{\pi({\bf Y}_{-i})}{\pi({\bf Y})}\right)$ as $N\rightarrow\infty.$

Substituting our estimators for each term of the theoretical KL divergence, we arrive at our overall KL-divergence based criterion,
\begin{align}
\label{eqn:klcrit1}
\widehat{\mathcal{D}}_i &=\begin{cases}
\frac{1}{N}\sum_{k=1}^N\log\left(f(y_i\vert\Theta^{(k)})\right) + \log\left(\frac{1}{N}\sum_{k=1}^N\left[f(y_i\vert\Theta^{(k)})\right]^{-1}\right) \text{ if } n_{j,(i)}^{(k)}-1\geq\minbot,\forall j,k\\
\infty  \text{ otherwise }
\end{cases}
\end{align}
where $f(y_i\vert\Theta^{(k)})=\frac{1}{\sqrt{2\pi}\sigma^{(k)}}\exp\left(-\frac{1}{2}\left(\frac{y_i-\sum_{j=1}^m\mu_{j,(i)}^{(k)}}{\sigma^{(k)}}\right)^2\right),$  $\mu^{(k)}_{j,(i)}$ is the terminal node parameter in tree $j$ from posterior sample $k$ to which observation $i$ maps, and $n_{j,(i)}^{(k)}$ is the number of obsevations mapping to the terminal node to which $y_i$ belongs in tree $j$ and posterior sample $k.$

\subsection{Conditional Predictive Ordinate Diagnostic}

Alternatively, %since $D_{KL}\geq 0,$ we know that the divergence is minimized when both terms in (\ref{eqn:kldiv}) are 
one can show that $D_{KL}$ can be rewritten as
\begin{align*}
	D_{KL} &=  \pi(y_i\vert{\bf Y}_{-i})^{-1}\int_\Theta log(f(y_i\vert\Theta))f(y_i\vert\Theta)\pi(\Theta\vert{\bf Y}_{-i}) + \pi(y_i\vert{\bf Y}_{-i})^{-1}\\
	&= CPO^{-1}\times \int_\Theta log(f(y_i\vert\Theta))f(y_i\vert\Theta)\pi(\Theta\vert{\bf Y}_{-i}) + log(CPO^{-1}).
\end{align*}
Note that the integrand $log(f(y_i\vert\Theta))f(y_i\vert\Theta)\rightarrow 0$ when the squared residual $e_i^2=(y_i-\mu)^2$ grows large.  This suggests the inverse CPO term in the KL-divergence is the important term to consider in identifying problematic observations, and as mentioned earlier, the inverse CPO has seen much use for exactly this purpose.
%minimized.  Note that the second term in (\ref{eqn:kldiv}) is constant for a given dataset and free of $\Theta.$  Since we do not directly compare values of $D_{KL}$ for different held-out observations (see Section 3.3), we need not concern ourselves with the second term and instead can concern ourselves only with small values of the first term.  That is, for some $\Theta^\dagger$ selected by the modeler such that $y_i-E[y_i\vert \Theta^\dagger]=k\sigma$ where $k$ is chosen by the modeler, then interest lies in how much larger the fitted model divergence is from this reference divergence level, i.e. $D_{KL}(\pi(\Theta\vert {\bf Y})\vert\vert\pi(\Theta\vert{\bf Y}_{-i}))-D_{KL}(\pi(\Theta^\dagger\vert {\bf Y})\vert\vert\pi(\Theta^\dagger\vert{\bf Y}_{-i})),$ where the terms involving the normalization constants, $\log(\pi({\bf Y}_{-i})/\pi({\bf Y})),$ cancel.  
%This suggests the diagnostic $\vert \int_{\boldmath{\Theta}}\log\left(f(y_i\vert\boldmath{\Theta})\right)\pi(\boldmath{\Theta}\vert{\bf Y})d\boldmath{\Theta} \vert,$ which can be approximated using the full-data posterior samples $\Theta^{(k)}\sim\pi(\Theta\vert {\bf Y})$ as
This motivates our third diagnostic, which approximates the (log) inverse CPO using the full-data posterior samples $\Theta^{(k)}\sim\pi(\Theta\vert {\bf Y})$ as in Proposition 0,
\begin{align}
\label{eqn:klcrit2}
\widetilde{\mathcal{D}}_i &=\begin{cases}
\log \left( \frac{1}{N}\sum_{k=1}^N f(y_i\vert\Theta^{(k)})^{-1}\right)  \text{ if } n_{j,(i)}^{(k)}-1\geq\minbot,\forall j,k\\
\infty  \text{ otherwise },
\end{cases}
\end{align}
and comparing $\widetilde{\mathcal{D}}_i$ to a reference level as outlined in the next section. %(\ref{eqn:klcrit2}) evaluated using $\Theta^\dagger.$
%The advantage of (\ref{eqn:klcrit2}) over (\ref{eqn:klcrit1}) is the reduced computations and, in particular, more stable computations since we need not take the sum of exponential terms seen in the second term of (\ref{eqn:klcrit1}).  In practice, we also find that (\ref{eqn:klcrit2}) more easily lends itself to a simple decision rule for detecting problematic observations based on the  choice of $k$ as described.

Overall, the advantage of these KL-divergence based diagnostics is that they tell us something about the sensitivity of the entire posterior distribution whereas the tree-based Cook's distance diagnostic (\ref{eqn:singletreecooks}) only tells us about the sensitivity of the mean function.  Nonetheless, we again see that (\ref{eqn:klcrit1}) and (\ref{eqn:klcrit2}) also exhibit inflationary behavior when we get into the degenerate situation of $n_j$ small, as determined by the minimum number of observations per terminal node parameter, $\minbot,$ which usually has a default value of $\minbot=5$ in most BART implementations.  The interpretation of these diagnostics is then clear: $\widehat{\mathcal{D}}_i$ and $\widetilde{\mathcal{D}_i}$ are large in the non-degenerate case when observation $i$ is far away from the other observations in its terminal node (since the density of $y_i$ will be small), or infinite in the degenerate case.

\subsection{Detecting Influential Observations}
In order to apply the above diagnostics, one requires a rule that will flag observations as potentially problematic.  Since we are operating in the Bayesian realm, a simple approach would be to take high-quantile values of the posterior samples of the diagnostics, such as the 97.5\% and 99\% quantiles, and use these as decision boundaries for detecting influentials.  However, this is approach is less than ideal since even in the case where there are no influential observations in a dataset, this approach will nonetheless flag 2.5\% or 1\% of observations as being problematic.

As motivated by the discussion for the CPO criterion (\ref{eqn:klcrit2}), we prefer the following alternative based on the notion of how large a residual would have to be in order to be considered problematic.  For Gaussian data, a residual that is $k=2$ standard deviations away would likely be the most conservative level most analysts would use to flag influentials, and $k=3$ standard deviations might be a more typical choice.  This implies substituting $e_i=2\sigma$ or $e_i=3\sigma$ in the diagnostics $D_i, \widehat{\mathcal{D}}_i$ and $\widetilde{\mathcal{D}}_i$ respectively.  For $D_i,$ one additionally needs to impute values for the tree complexity and node purity terms.  One approach would be to substitute posterior averages from the fitted model.  Alternatively, we can impute values based on the priors; this suggests $1/8$ for the tree complexity term and since the default value of $\minbot$ is typically $5$, this suggests $5/16$ for the node purity term.  Calibrating the decision rules in this way is intuitive and interpretable for the practitioner, and in our applications appears to work quite well.  See, for instance, Figures 3-6 that apply this rule using $2\sigma$ and $3\sigma$ cutoffs in Section \ref{section:examples}.  %An exception is the $\widehat{\mathcal{D}}_i$ statistic, which generally shows little difference between $2\sigma$ and $3\sigma$ cutoffs, lending a preference for $\widetilde{\mathcal{D}}_i$ in practice.

%% file: importance.tex
While one could use the proposed diagnostics to detect problematic observations and then refit the model with such observations removed from the dataset, for Bayesian models implemented using MCMC sampling algorithms (such as BART), this is a computationally wasteful approach.  Instead, \cite{bradlow:zaslavsky:1997} propose to  estimate functions of interest, $g(\boldmath{\Theta}),$ using importance sampling as
\[
E[g(\Theta)\vert {\bf Y}_{-i}]=\int_{\Theta} g(\Theta)\frac{\pi(\Theta\vert {\bf Y}_{-i})}{\pi(\Theta\vert {\bf Y})}\pi(\Theta\vert{\bf Y})d\Theta.
\]
Let $w_{(i)}^{(k)}=\frac{\pi(\boldmath{\Theta}=\boldmath{\Theta}^{(k)}\vert{\bf Y}_{-i})}{\pi(\boldmath{\Theta}=\boldmath{\Theta}^{(k)}\vert{\bf Y})}\propto\frac{f({\bf Y}_{-i}\vert \boldmath{\Theta}=\boldmath{\Theta}^{(k)})}{f({\bf Y}\vert \boldmath{\Theta}=\boldmath{\Theta}^{(k)})}$ be the importance sampling weights of interest when observation $i$ is to be dropped and  $\boldmath{\Theta}^{(k)}\sim\pi(\boldmath{\Theta}\vert{\bf Y}).$  Then,
\[
E[g(\Theta)\vert {\bf Y}_{-i}]\approx \frac{\sum_{k=1}^N w_{(i)}^{(k)}g(\Theta^{(k)})}{\sum_{k=1}^N w_{(i)}^{(k)}},
\]
where the renormalization in the denominator removes the dependence on the proportionality constant $\pi({\bf Y})/\pi({\bf Y}_{-i}).$  Intuitively, this importance sampling approach adjusts our posterior samples used in predicting  $g(\boldmath{\Theta})$ as if we had instead sampled from $\boldmath{\Theta}\vert {\bf Y}_{-i}.$  The weights also have a clear connection to the KL-divergence diagnostic proposed in Section \ref{section:diagnostics}, with the difference being that the diagnostic is based on the log density ratio whereas the weights are calculated on the density ratio scale.  However, it turns out that direct application of these weights to posterior quantities of interest does not behave well due to the high-dimensional parameter space of treed models, particularly the richer models such as BART.  This is because in a high-dimesional parameter space, the localized parameters affected by the problematic observation tend to be uncorrelated with poor draws for the rest of the high-dimensional parameter, and so downweighting entire posterior realizations from such high-dimensional parameter spaces tends to remove good samples for the rest of the parameter space.  This problem with the ``global reweighting'' scheme can only be overcome by collecting extremely large numbers of posterior samples, which is computationally prohibitive.  Fortunately, careful investigation of the situation in the prediction setting yields an effective reweighting scheme.

\subsection{Re-weighting Bayesian Tree Predictions}

As we are typically interested in prediction, we will focus on a reweighting scheme for posterior quantities involving the terminal node parameters. The conditional independence structure of Bayesian trees allow us to simplify the calculation of the importance sampling weights while increasing their effectiveness in practice.  Suppose the observation to be removed, $y_i$, belongs to terminal node $\eta_j$ with mean parameter $\mu_j$ in the current tree defined by $\boldmath{\Theta}.$  Furthermore, let $P_j$ be the set of internal nodes with associated split rules that define the path from $\eta_j$ back to the root node, and let $\widetilde{\boldmath{\Theta}}=\boldmath{\Theta}\setminus P_j,\mu_j$ represent the remaining tree parameters.  Note that $P_j$ implicitly defines a hyperrectangle in the input space that maps to terminal node $\eta_j$ with associated prediction $\mu_j.$   Then we have the following.

{\bf Proposition 1:} Consider functions $g(\boldmath{\Theta})\equiv g(\mu_j),$ such as predictions involving only terminal node $\eta_j.$  Then, the weights are given by $w^{(k)}_{(i)}\propto f^{-1}(y_i\vert\mu_j,P_j,\sigma^2)\I(\vert\eta_j\vert-1\geq\minbot)$ where $\vert\eta_j\vert$ is the number of observations from the full dataset ${\bf Y}$ that map to terminal node $\eta_j$ and $n_0$ is the minimum number of observations allowed per terminal node.  Similarly, consider functions $g(\boldmath{\Theta})\equiv g(\mu_l), l\neq j$ such as predictions {\em not} involving terminal node $\eta_j$.  Then the weights are $w^{(k)}_{(i)}= 1.$

Note that this result still holds in the case that $\minbot=0.$  In words, Proposition 1 says that when we hold-out $y_i$, the weights for predictions involving the subregion of the input space defined by $P_j$ involves re-weighting the predictions by the inverse density in $y_i$ if the node would have been valid with the case deleted, otherwise the prediction receives zero weight.  Meanwhile, weights for predictions involving terminal nodes other than $\eta_j$ effectively receive a weight of 1, indicating no ill effect of the case deletion and lending the interpretation that influence in Bayesian tree models has a local effect in terms of prediction.

The conditional independence structure of Bayesian trees allows this idea to be extended to functionals of other tree parameters, or more than single-case deletion, although the practical calculation may be come unwieldy as the factorization of the tree becomes more complex.

\subsection{Re-weighting BART Predictions}
We can extend the idea of reweighting to draws from the additive tree model of BART.  Consider using BART's $m$-tree ensemble to predict at some new input setting $x.$  The added complexity in this situation arises from the possibility that all $m$ terminal nodes from the $m$ trees that will be used to predict the response at $x$ may have full, partial or no dependence on the problematic observation $y_i.$  That is, all $m$ terminal nodes may have contained $y_i$, or some subset of the $m$ terminal nodes may have contained $y_i$, or perhaps none.  Proposition 2 describes the weighting scheme in this case.

{\bf Proposition 2:} Consider functions $g(\boldmath{\Theta})=g(\sum_{j=1}^m\mu_j(x))$, such as predictions involving only the $m$ terminal nodes, $\boldsymbol{\eta}_x,$ to which input $x$ maps.  Suppose $y_i$ maps to the $m$ terminal nodes $\boldsymbol{\eta}_{x_i}.$
Let $\boldsymbol{\eta}_a=\boldsymbol{\eta}_{x_i}\cup \boldsymbol{\eta}_x.$  Then, if at least one of the terminal nodes in $\boldsymbol{\eta}_x$ is in $\boldsymbol{\eta}_{x_i}$ the weight is $w^{(k)}_{(i)}(x)\propto \frac{1}{f(y_i\vert\boldsymbol{\mu}_a,\boldsymbol{\eta}_a,{\bf P}_a,\sigma^2)}\prod_{\eta*\in \boldsymbol{\eta}_{x_i}}\I(\vert\eta^*\vert-1\geq \minbot).$   Analogously, for predictions at $x$ which do not map to any terminal node in $\boldsymbol{\eta}_{x_i},$ the corresponding weight is $1.$

Proposition 2 essentially says that predictions involving any subset of the terminal nodes to which $x_i$ maps will be reweighted, and the weights are essentially the same except for the indicator function verifying the $\minbot$  constraint.  That is, the union of the rectangular regions defined by the terminal nodes to which $x_i$ maps will be reweighted when predicting at a new $x$ that lies somewhere in this union.  This means calculating the weights is relatively more complex than the single-tree case described earlier, and it also suggests that the weighting will often be inefficient much as the original method of \cite{bradlow:zaslavsky:1997} was when applied to the single-tree case.  This is motivated by the fact that BART prefers shallow trees, and so the union of regions involving the $x_i$ across all $m$ trees may in fact be quite large.

It is tempting, then, to consider a more localized variant -- a weighting scheme that only involves predictions that fall in the intersection of regions defined by the terminal nodes to which $x_i$ maps.  In fact, such an approach can again be supported by recalling that the BART likelihood involving a sum-of-trees mean function can, conditionally, be equivalently described by a single ``super-tree'' mean function \citep{horiguchi:etal:2021}, that is
\begin{align*}
\pi(\boldmath{\Theta}\vert {\bf Y})\propto f\left({\bf Y}\vert ({\bf T}^{(1)},{\bf M}^{(1)}),\ldots,({\bf T}^{(m)},{\bf M}^{(m)}),\sigma^2\right)\prod_{k=1}^m\pi\left(({\bf T}^{(k)},{\bf M}^{(k)})\right)\pi\left(\sigma^2\right)\\
=f\left({\bf Y}\vert \mathcal{S},\sigma^2\right)\prod_{k=1}^m\pi\left(({\bf T}^{(k)},{\bf M}^{(k)})\right)\pi\left(\sigma^2\right),
\end{align*}
where $\mathcal{S}$ represents the analogous super-tree representation, i.e. $g(x;\mathcal{S})\equiv\sum_{k=1}^mg(x;({\bf T}^{(k)},{\bf M}^{(k)})).$  Note that the prior remains the same, even though we reinterpret the likelihood's sum-of-trees as a new, equivalent, single super-tree.  Suppose again that $y_i$ is the problematic observation, observed at input $x_i$ and let $\eta^\mathcal{S}_l$ be the terminal node in $\mathcal{S}$ to which $x_i$ maps, noting that there is a single unique such terminal node in $\mathcal{S}.$  Let $\mathbb{X}$ represent the hyperrectangle defined by $\eta^\mathcal{S}_l$.  Then we have the following.

{\bf Proposition 3:}
Let $x$ be a prediction input of interest.  Let $\mathbb{X}_j$ be the hyperrectangles in each tree $j=1,\ldots,m$ of the BART ensemble such that $x\in \mathbb{X}_j,\forall j.$  Let $\mathbb{X}=\cap_{j=1}^m\mathbb{X}_j$ be the hyperrectangle defined as the intersection of all the $\mathbb{X}_j$'s, which corresponds to the supertree terminal node $\eta^\mathcal{S}_l$ to which $x$ belongs.  Supppose also that the input $x_i$ for influential observation $y_i$ also maps to $\eta^\mathcal{S}_l.$  Then to predict the response $y(x)$ for all $x\in\mathbb{X},$ the weights are
\[
w^{(k)}_{(i)}(x)=
\begin{cases}
\frac{1}{p(y_i\vert \mu^{\mathcal{S},(k)}_l,\eta^{\mathcal{S},(k)}_{l},P^{\mathcal{S},(k)}_{l},\sigma^2)} & \text{ if } \vert \eta^{(k)}_{jl}\vert-1 \geq m \text{ for all } k=1,\ldots,m\\
0 & \text{ otherwise }
\end{cases}
\]
where $\eta^{(k)}_{jl}$ is the $l$th terminal node in tree $j$ to which $x_i$ maps in the original sum-of-trees representation. 

Note in this version of the weighting scheme, the observation $x_i$ only maps to a single terminal node in the supertree representation, and this node corresponds to the intersection of rectangular regions defined by all of the $m$ terminal nodes involving $x_i$ in the original sum-of-trees representation.  As such, Proposition 3 defines a more localized weighting scheme, and is also easier to manage from an implementation perspective.

\subsubsection{A union of intersections}
The practical implementation of Proposition 3 results in a different localized region, say $\mathbb{X}^{(k)},$ for each of the $k=1,\ldots,N$ posterior realizations.  This makes predictions more computationally expensive. A practical alternative is to take some sort of ``average'' localized region as the single region to reweight, simplifying posterior prediction calculations.  A natural choice is the union of the individual regions, say $\overline{\mathbb{X}}=\cup_{k=1}^N\mathbb{X}^{(k)}.$  Not only does this simplify the calculation of posterior predictions, it also results in only requiring a single region $\overline{\mathbb{X}}$ to be saved from model training, reducing the amount of memory required to store the model.  Since each $\mathbb{X}^{(k)}$ is itself a region defined by the intersection resulting from the supertree, we refer to this method as \texttt{union-int}.

\subsubsection{An L1 distance alternative}
Since the reweighting region defined by $\overline{\mathbb{X}}$ is simply a hyperrectangle, it is tempting to consider a less involved approach. One possibility is, upon identifying a problematic observation, to take an L1 region around this point, say $\widetilde{\mathbb{X}},$ as the region to be reweighted.  That is, take $\widetilde{\mathbb{X}}=\lbrace x: \vert\vert x_v-x_{iv}\vert\vert_{1}<\delta\ \forall\ v=1,\ldots,d \rbrace$ for some well-chosen scalar constant $\delta.$
We refer to this method as $\ell_1,$ and briefly consider this empirical alternative in Section 5.2.  However, it turns out to not be practically useful as choosing a good $\delta$ is itself an expensive optimization.

%% file: examples.tex
\subsection{Motivating Examples}
To motivate our influence metrics, we start with a simple 1-dimensional and 2-dimensional test function.  The 1D function is cubic, having an input domain $[0,1]$ and response values calculated as $f(x)=8*(x-0.5)^3.$  The $n=100$ observations are generated as $y=f(x)+\epsilon,\ \epsilon\sim\text{N}(0,0.05^2).$  The 2D function is taken to be the popular Branin function.  The Branin function is a smoothly varying response surface computed over the 2-dimensional domain $x\in [0,1]^2$ as $$f(x)=\frac{1}{51.95}\left[(\bar{x}_2-\frac{5.1\bar{x}_2}{4\pi^2}+\frac{5\bar{x}_1}{\pi}-6)+\left(10-\frac{10}{8\pi}cos(\bar{x}_1)-44.81\right)\right]$$ where $\bar{x}_1=15x_1-5,\bar{x}_2=15x_2$ \citep{picheny2013benchmark}. The function exhibits steep slopes in some regions of the input space, particularly along the edges of the domain.  The $n=500$ observations were generated as $y=f(x)+\epsilon,\ \epsilon\sim\text{N}(0,0.05^2).$  For both of these simple functions we fit the BART model using the default options, in particular $m=200$ trees, $k=2$, $\texttt{numcut}=100$, and a minimum of $\minbot=5$ observations per terminal node.\\

\subsubsection{Diagnostics}
First, we consider the influence diagnostics and investigate two scenarios: no influential observations and influential observations with a residual of 3 s.d.  We refer to Cook's distance (equation \ref{eqn:singletreecooks}) as \texttt{cooks}, the KL-divergence metric $\widehat{\mathcal{D}}_i$ (equation 5) as \texttt{KL} and the log inverse CPO metric $\widetilde{\mathcal{D}}_i$ (equation 6) as \texttt{CPO}. As a reference, we compute \texttt{cooks} and \texttt{CPO} (\ref{eqn:kldiv}) by plugging in $2$ and $3$ s.d. residuals with posterior mean estimates of other relevant quantities to serve as a gauge of severity of the calculated diagnostics of each observation.  For \texttt{KL} we use the estimated posterior $97.5$th and $99.5$th quantiles.  For the scenarios where influential observations were constructed, two such observations were formed: influential observation \#1 at the center of the regression domain ($0.5$ and $(0.5,0.5)$ for cubic and Branin respectively), and one at an edge of the domain ($1.0$ and $(0.0,1.0)$ for cubic and Branin respectively).\\

\begin{figure}
\centering
\includegraphics[scale=1]{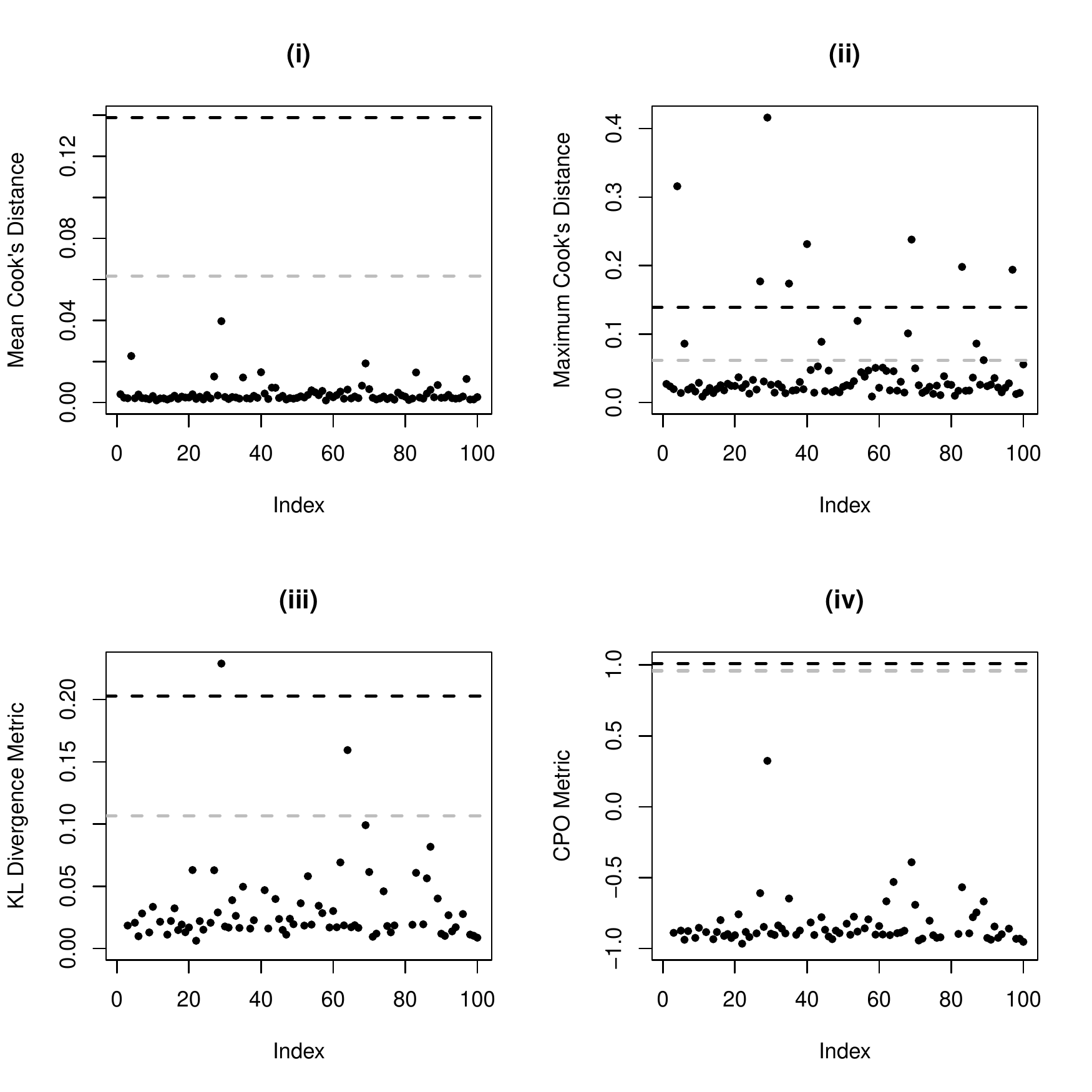}
\caption{Influence diagnostics for the 1D cubic test function with no constructed influentials.  Panel (i) displays the mean \texttt{cooks} diagnostic; (ii) displays the maximum \texttt{cooks} diagnostic; (iii) displays the \texttt{KL} divergence diagnostic (excluding infinities); and (iv) shows the \texttt{CPO} diagnostic (excluding infinities). Grey dashed line denotes the $2\sigma$ cut-off while the black dashed line denotes the $3\sigma$ cut-off.%location of observations that the KL-divergence diagnostic evaluated to infinity due to too few observations remaining in those observations' terminal nodes.
}
\label{fig:cubic}
\end{figure}

{\bf 1D Cubic}\\
With no influential observations, the results of computing the three discussed diagnostics are shown in Figure \ref{fig:cubic}.  This figure shows that mean \texttt{cooks} and \texttt{KL} diagnostics do a good job when there is nothing to detect.  The maximum \texttt{cooks} and \texttt{KL} diagnostics appear overly sensitive as some observations are suggested as problematic.  For \texttt{KL} this is not surprising as there will always be some diagnostic values falling above the empirical $97.5$th and $99.5$th quantiles.  Note that some observations also evaluate to infinity for the \texttt{KL} and \texttt{CPO} diagnostics.  These observations violate the $\minbot$ constraint of the model, and the locations of these observations tend to occur at the edges of the prediction domain, or where there are gaps in the data (not shown).\\

\begin{figure}
\centering
\includegraphics[scale=1]{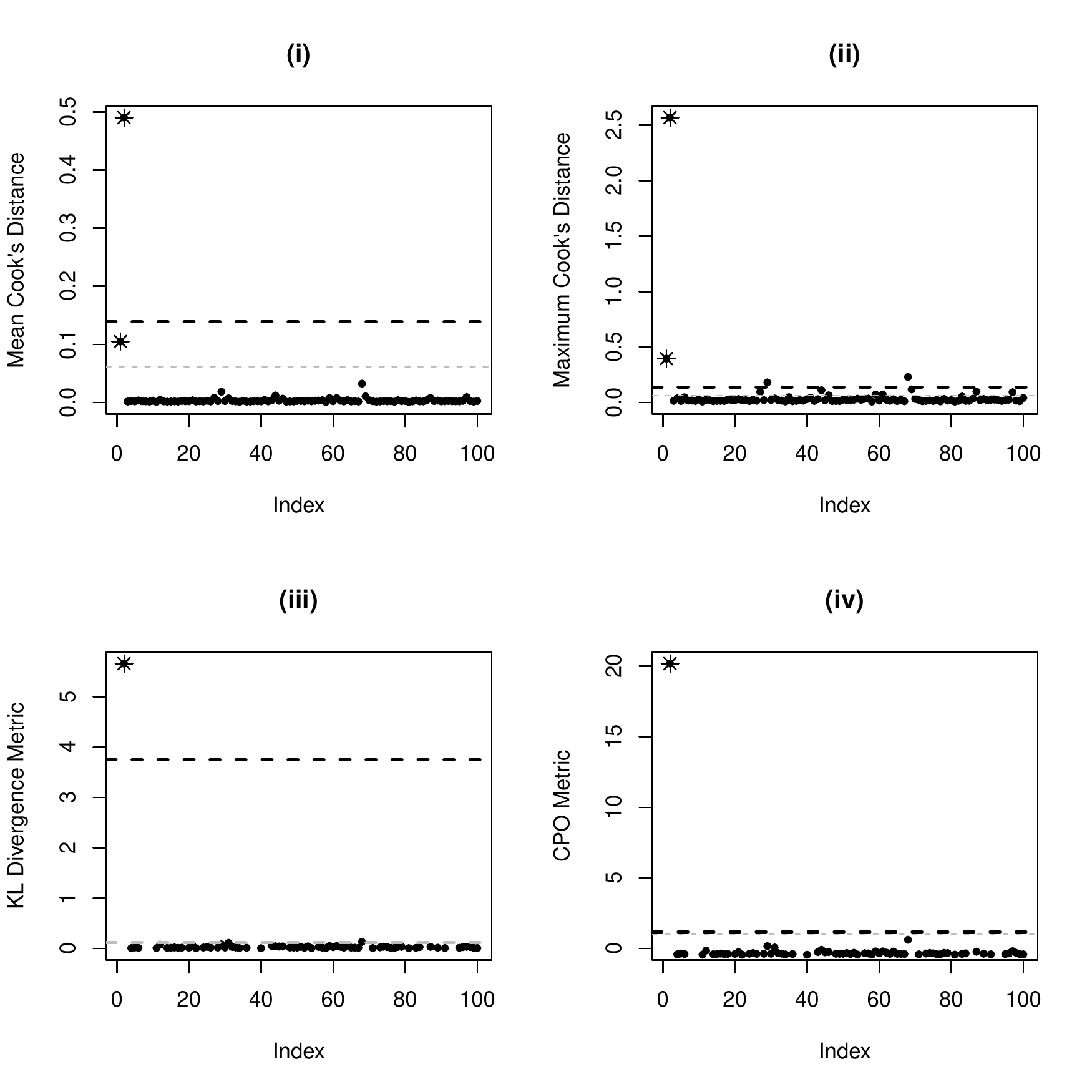}
\caption{Influence diagnostics for the 1D cubic test function with influentials at $x=0.5$ and $x=1$.  Panel (i) displays the mean \texttt{cooks} diagnostic; (ii) displays the maximum \texttt{cooks} diagnostic; (iii) displays the \texttt{KL} diagnostic (excluding infinities); and (iv) shows the \texttt{CPO} diagnostic (excluding infinities).  The true influential observations are denoted by $\Asterisk.$  Grey dashed line denotes the $2\sigma$ cut-off while the black dashed line denotes the $3\sigma$ cut-off. %shows the location of observations that the KL-divergence diagnostic evaluated to infinity due to too few observations remaining in that observations terminal nodes.
}
\label{fig:cubic3}
\end{figure}

The results once we add in the influential observations are shown in Figure \ref{fig:cubic3}.  We can see that both the mean \texttt{cooks}, \texttt{KL} and \texttt{CPO} diagnostics are able to pick up the influential observations accurately (the \texttt{KL} and \texttt{CPO} for the observation at $x=1$ in fact evaluates to infinity, so is not shown in panels (iii) and (iv)).  We also note that observation \#2, located at $x=1$, exerts greater influence than the observation at $x=0.5$, as one would expect.  The maximum \texttt{cooks} diagnostic also easily detects the two influentials, but also suggests a few other observations might be problematic when they are not.  Similarly, \texttt{KL} easily detects the influentials but again indicates its propensity to suggest problematic observations where there are none.  Note again that the observations for which the \texttt{KL} and \texttt{CPO} diagnostics evaluate to infinity tend to occur at the edges of the domain, where the $\minbot$ constraint is most likely to be violated.\\

\begin{figure}
\centering
\includegraphics[scale=1]{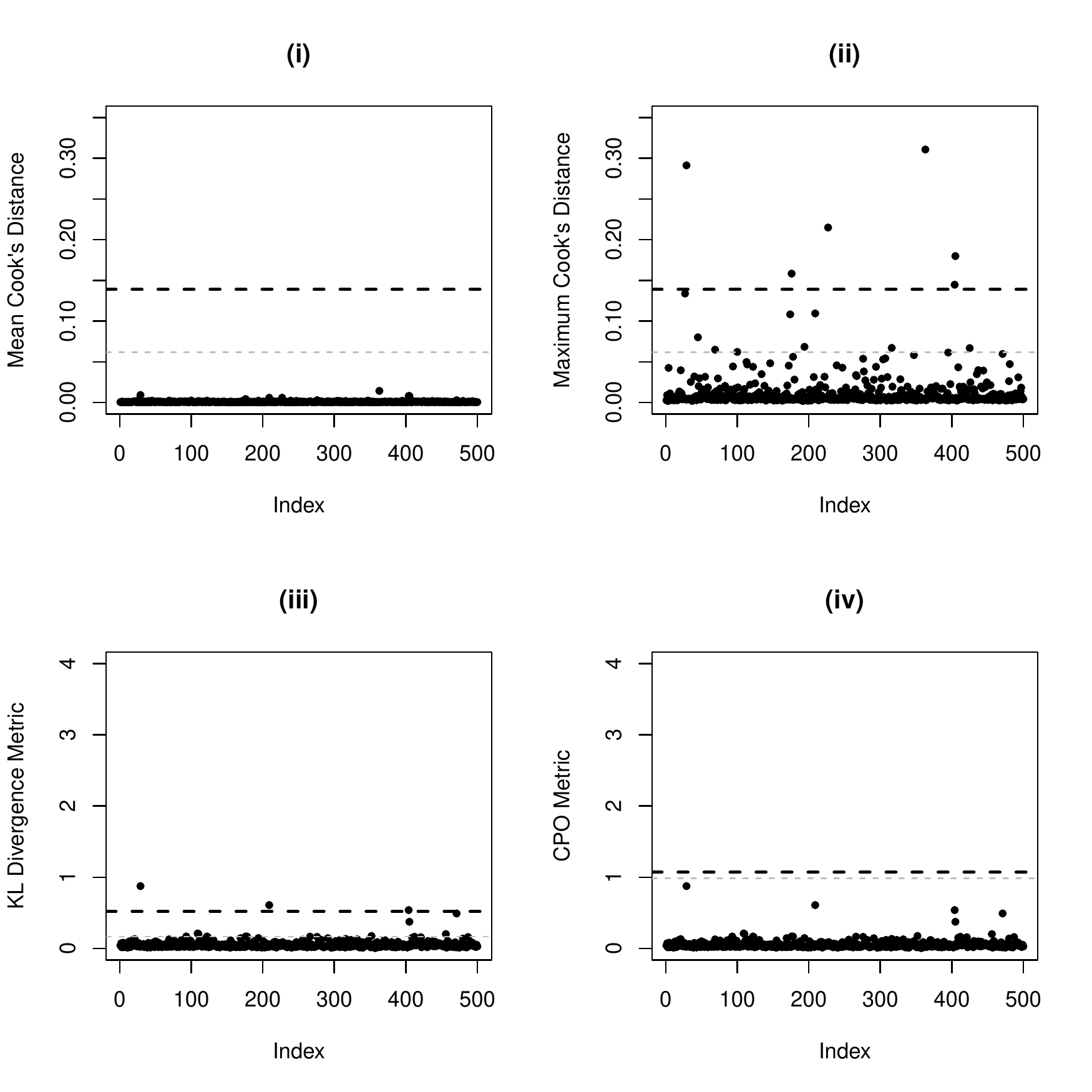}
\caption{Influence diagnostics for Branin test function with no constructed influentials.  Panel (i) displays the mean \texttt{cooks} diagnostic; (ii) displays the maximum \texttt{cooks} diagnostic; (iii) displays the \texttt{KL} diagnostic (excluding infinities); and (iv) displays the \texttt{CPO} diagnostic (excluding infinities). Grey dashed line denotes the $2\sigma$ cut-off while the black dashed line denotes the $3\sigma$ cut-off. %shows the location of observations that the KL-divergence diagnostic evaluated to infinity due to too few observations remaining in that observations terminal nodes.
}
\label{fig:branin}
\end{figure}

{\bf Branin}\\
The results for no influential observations for the Branin function are shown in Figure \ref{fig:branin}.  Here we see that, as expected, the mean \texttt{cooks} and \texttt{CPO} diagnostics do not suggest any problematic observations.  The maximum \texttt{cooks} and \texttt{KL} diagnostics again suggest potentially problematic observations, which might indicate that these diagnostics are overly sensitive.  As before, the \texttt{KL} and \texttt{CPO} diagnostics do not plot any observations whose criterion evaluated to infinity.  In fact, 8 observations in this example did evaluate to infinity, indicating that the $\minbot$ limit was violated once those observations were held out from their respective terminal nodes.  These observations %are plotted in $x$-space in Figure \ref{fig:branin}(iv), where we note that these observations are 
generally occured at the edges of the domain and/or in regions where the response is changing rapidly.  These are scenarios where it is known that the quality of BART's fit can suffer, and it is interesting that these diagnostics (and possibly the maximum \texttt{cooks} diagnostic) are able to detect such issues.\\

\begin{figure}
\centering
\includegraphics[scale=1]{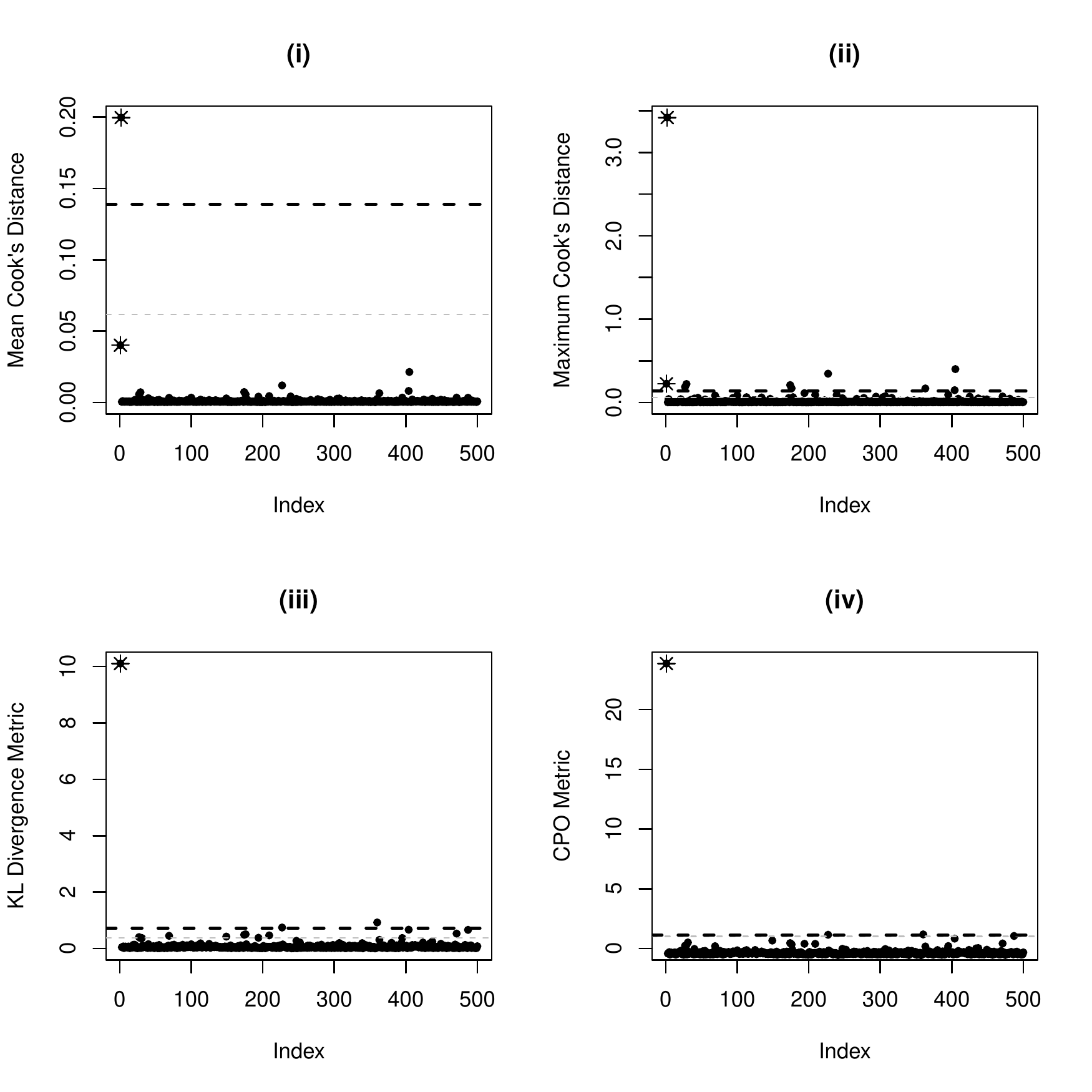}
\caption{Influence diagnostics for Branin test function with influentials at $x=(0.5,0.5)$ and $x=(0,1)$.  Panel (i) displays the mean \texttt{cooks} diagnostic; (ii) displays the maximum \texttt{cooks} diagnostic; (iii) displays the \texttt{KL} diagnostic (excluding infinities); and (iv) displays the \texttt{CPO} diagnostic (excluding infinities).  The true influential observations are denoted by $\Asterisk.$  Grey dashed line denotes the $2\sigma$ cut-off while the black dashed line denotes the $3\sigma$ cut-off. %Panel (iv) shows the location of observations that the KL-divergence diagnostic evaluated to infinity due to too few observations remaining in that observations terminal nodes.
}
\label{fig:branin3}
\end{figure}

Adding in influential observations results in the diagnostic outputs shown in Figure \ref{fig:branin3}.  Here we see that the mean \texttt{cooks} diagnostic is able to pick up the influential at (0,1) easily and also suggests a potential problem with the influential at (0.5,0.5), although some observations around index 200 and index 400 give similar diagnostic values.  The maximum \texttt{cooks} diagnostic easily detects the influential at (0,1), but also flags a few observations around index 200 and index 400 while barely detecting the influential at (0.5,0.5).  These plots suggest that while the \texttt{cooks} diagnostics can be useful, they may also suffer from higher than desired false positive and false negative errors.  The \texttt{KL} diagnostic has fairly good performance but also suggests a few observations that might be flagged as problematic when they are not. The \texttt{CPO} diagnostic appears to be the most powerful of the four - it easily, and strongly detects the two influential observations and succesfully ignores the observations that were not influential.   And, as an added feature, this diagnostic again detected observations along the perimeter of the domain that evaluate to infinity (not shown).\\%, as shown in Figure \ref{fig:branin3}(iv).\\

\subsubsection{Reweighting Predictions}
Given the successful identification of problematically influential observations, an existing fit to our respective test functions can be reweighted to alleviate the impact of the influentials.  We apply all three methods described in Section \ref{section:importance}, including the method of \cite{bradlow:zaslavsky:1997} which we refer to as \texttt{global} and the proposed methods of Propositions 1, 2 and 3 which were refer to as \texttt{union}, \texttt{int} and \texttt{union-int} respectively.

\begin{figure}
\centering
\includegraphics[scale=1]{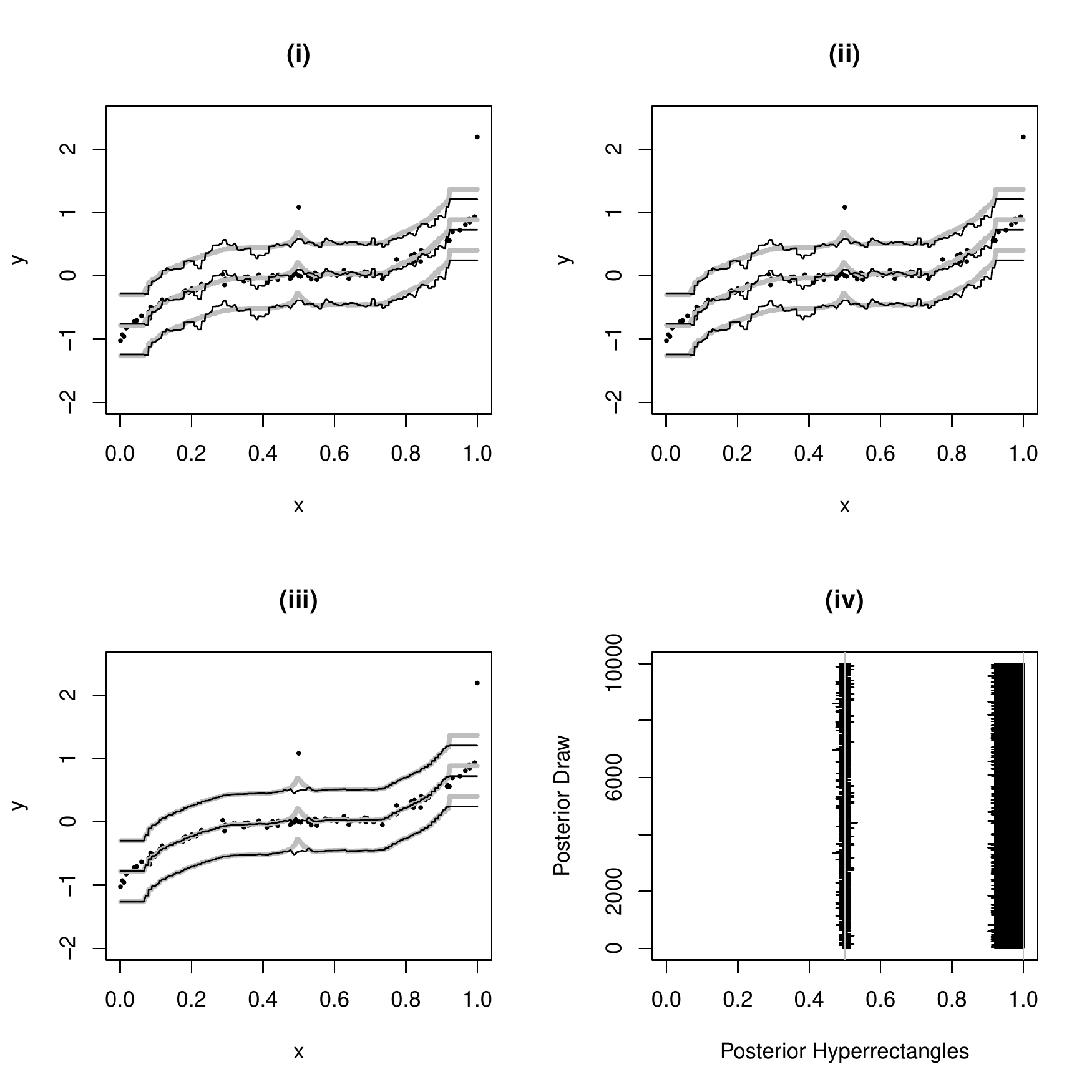}
\caption{Reweighting psoterior predictive distribution draws of fitted cubic test function with influential observations at $x=0.5$ and $x=1.$  The original BART fit (uncorrected) is shown as the light gray lines along with +/-2sd credible intervals, while corrected predictions and intervals are shown in black.  Panel (i) shows the \texttt{global} correction, (ii) shows the \texttt{union} correction and (iii) shows the \texttt{int} correction.  In panel (iv), the hyperrectangular regions to which the {\texttt int} correction is applied is shown for all 10K posterior draws.  In comparison, the \texttt{union} correction is applied to the entire [0,1] domain, resulting in the same performance as \texttt{global} in this example.}
\label{fig:cubicreweighted}
\end{figure}

The results for the cubic test function are shown in Figure \ref{fig:cubicreweighted}.  The original BART fit (light grey) demonstrates the local effect of the influential observations located at $x=0.5$ and $x=1$ respectively.  The three reweighting methods are summarized in Figure \ref{fig:cubicreweighted}(i)-(iii).  From this example we observe that \texttt{global} is the worst of the reweighting methods, noticeable affecting the quality of fit away from the influential observations.  The \texttt{union} method, in this case, matches \texttt{global}'s performance.  This somewhat counter-intuitive behavior arises from the fact that the union of hyperrectangles in this method ends up being the entire [0,1] input domain.  The \texttt{int} method demonstrates much better performance, having nearly identical model fit quality as the original BART posterior away from the influential observations while correcting for the influential observations in their respective localities.  These localities, defined by the intersection of hyperrectangles in this case, are shown over all 10K posterior draws in Figure \ref{fig:cubicreweighted}(iv).  Finally, the \texttt{union-int} provides the best performance by `collapsing' the posterior draws in Figure \ref{fig:cubicreweighted}(iv) while also being computationally cheaper to perform.

\begin{figure}
\centering
\includegraphics[scale=0.45]{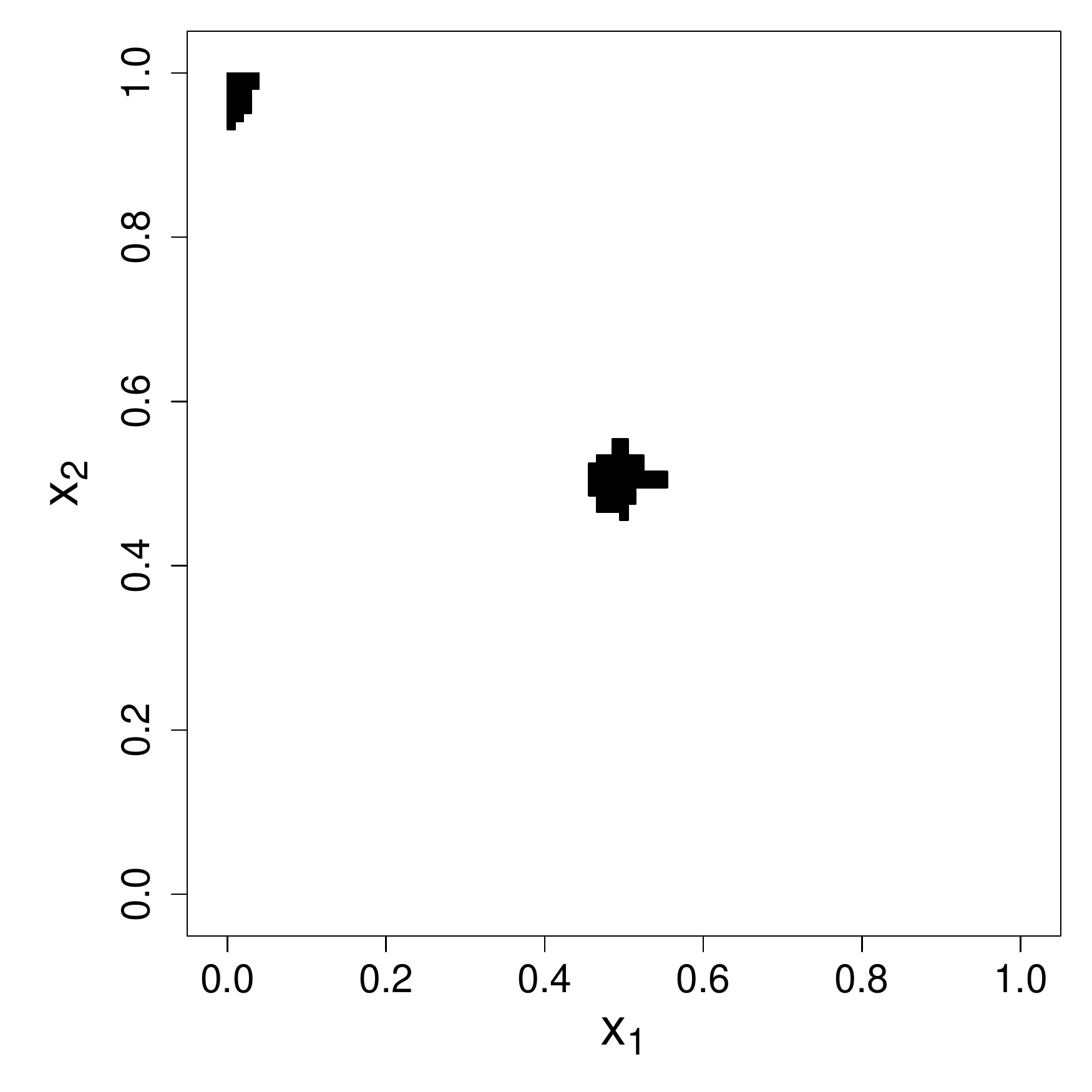}
\includegraphics[scale=0.45]{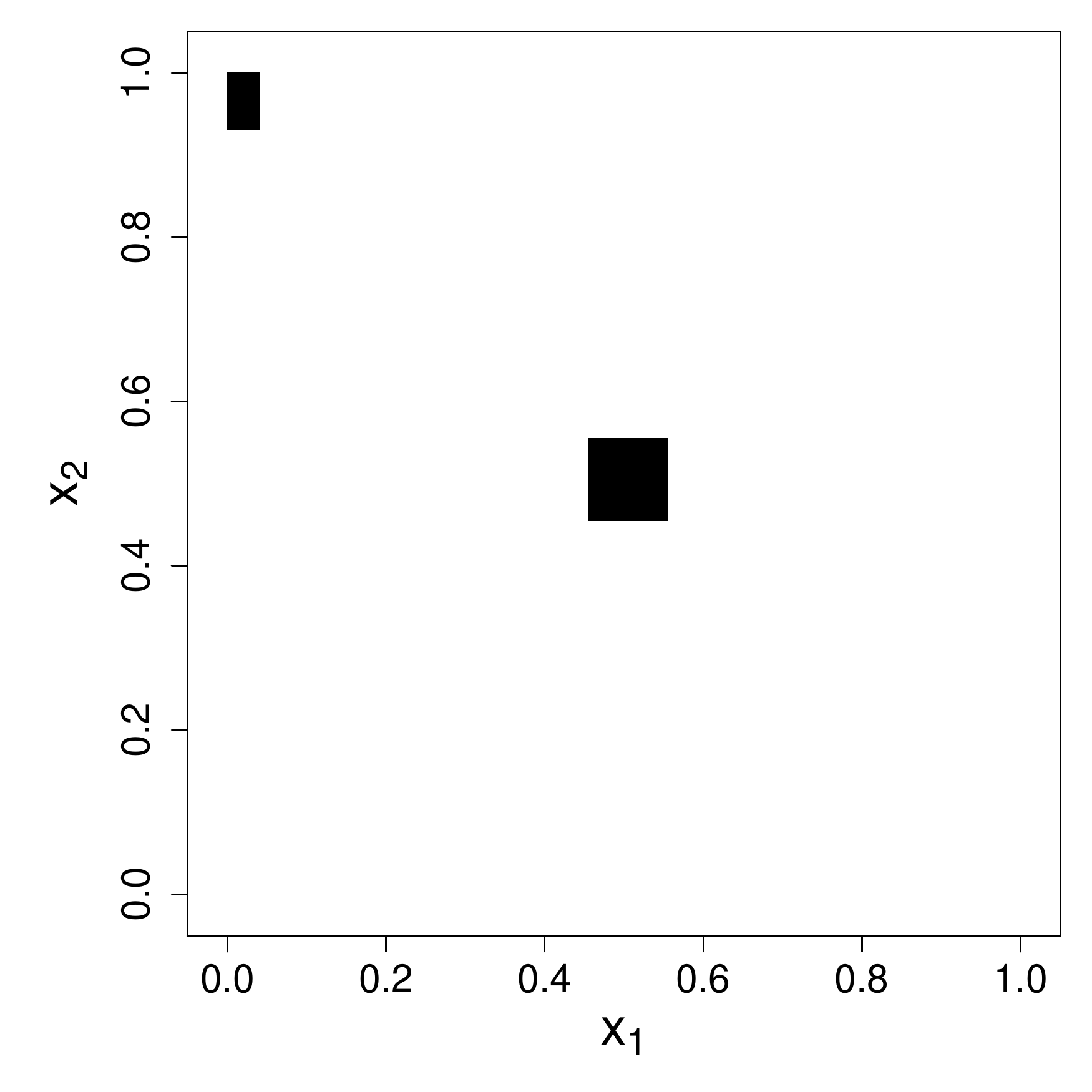}
\caption{Posterior hyperrectangle regions of influence for the Branin test function using the \texttt{int} method (left panel) and \texttt{union-int} method (right panel).  The true influential observations are located at input settings $x=(0.5,0.5)$ and $x=(0.0,1.0).$}
\label{fig:braninhyperrects}
\end{figure}

A similar behavior is seen for the Branin test function.  Table \ref{tab:branin} summarizes the performance by looking at in-sample and out-of-sample RMSE for the various BART predictors.  Similar to the cubic test function, we see that the \texttt{global} and \texttt{union} methods have equal performance since \texttt{union} again results in the union of hyperrectangles being the entire $[0,1]^2$ input domain.  Both methods introduce variance in the predictor that inflates the prediction error relative to BART fit without including the influentials, denoted as \texttt{oracle}.  Meanwhile, the \texttt{int} method again exhibits performance on par with the \texttt{oracle} BART fit by removing the influence of the outliers located at $x=(0.5,0.5)$ and $x=(0.0,1.0).$  The posterior intersection hyperrectangles detected by \texttt{int} are shown in Figure \ref{fig:braninhyperrects} (left panel), confirming that the reweighting procedure is being applied in appropriate regions of the input space.  Finally, the \texttt{union-int} provides slightly better performance than \texttt{int} by taking the regions shown in Figure \ref{fig:braninhyperrects} (right panel).

\begin{table}[h!]
  \begin{center}
    \caption{RMSE performance of BART predictors for the Branin test function.}
    \label{tab:branin}
    \begin{tabular}{|c|c|c|c|c|c|}
	\hline
      - & \texttt{oracle} & \texttt{global} & \texttt{union} & \texttt{int} & \texttt{union-int} \\
      \hline
      \hline
%      in-sample & 0.099 & 0.172 & 0.172 & 0.106\\
%      out-sample & 0.136 & 0.194 & 0.194 & 0.139\\
      in-sample & 0.0540 & 0.0904 & 0.0904 & 0.0564 & 0.0531\\
      out-sample & 0.1584 & 0.1595 & 0.1595 & 0.1528 & 0.1463\\
	  \hline
    \end{tabular}
  \end{center}
\end{table}

\subsection{Simulation Study}

For a broader persepctive on the performance of our detection and reweighting methods, we considered a simulation study using the 5-dimensional Friedman test function, defined as $f({\bf x})=10sin\left(\pi x_1x_2\right)+20\left(x_3-0.5\right)^2+10x_4+5x_5$  where the input domain is $x\in [0,1]^5.$  We consider a single influential observation at the centroid, $x_{infl}=(0.5,0.5,0.5,0.5,0.5)$ and the influential observation is generated using an offset of 5.  We also explore $m=1$ and $m=200$ settings reflecting single-tree and default BART models respectively, and vary the sample size as $n=50\ (m=1\ \text{only}),\ n=100$ and $n=500.$  All other settings, in particular $\minbot=5,$ were left at the BART defaults.  At each of these experimental settings, 100 replicate runs were performed by generating a new dataset, fitting BART, and then calculating the usual BART posterior prediction.  Performance was measured in terms of local and global prediction performance.  Global prediction was estimated by evaluating the prediction error at $n_p=5,000$ out-of-sample inputs drawn in $[0,1]^5$ while local prediction error considered $n_p=5,000$ out-of-sample inputs drawin in $[0.4,0.6]^5.$  

To generate the data with the outlier being influential enough to be detected and corrected, one can use the nice interpretation of (\ref{eqn:singletreecooks}) to motivate the offset to add to the influential observation.  Equation (\ref{eqn:singletreecooks}) allows one to ask {\em how many standard deviations away (say $k$) would an influential observation need to be to be as influential as an observation 2 standard deviations away when $m=\minbot$}?  The solution is given by the inequality $k>\sqrt{2^2\times\frac{\minbot}{\minbot-1}\times\frac{n_*}{(n_*-1)^2}}$ where we can take $n_*$ to be the typical number of observations in a terminal node.  Under the default tree prior we expect no more than 8 terminal nodes, so with a simulation study of $n=50-500$ a reasonable range for $n_*$ is $6-60.$  This results in $k$ ranging from $2.28 - 8.5$; we take $k=5.$  Finally, since we generate the data with $\sigma=1,$ a reasonable offset for our simulated influential observation is therefore $5.$

\begin{table}[h!]
\small
	\begin{center}
	\caption{Local RMSE performance in region around influential observation for Friedman simulation study.}
    \label{tab:friedstudy:a}
		\begin{tabular}{|c|c|c|c|c|c|c|}
		\hline
%		$m$ & $n$ & \texttt{default} & \texttt{oracle} & \texttt{global} & \texttt{union} & \texttt{int} & \texttt{union-int} & $\ell_1$\\
%		\hline
%		\hline
%		1   & 500 & 2.91 & 2.74 & 2.83 & 2.81 & 2.81 & 2.83 & 2.86\\
%		1   & 100 & 2.74 & 2.47 & 2.46 & 2.48 & 2.48 & 2.46 & 2.57\\
%		1   & 50  & 2.45 & 2.29 & 2.27 & 2.39 & 2.39 & 2.27 & 2.39\\
%		200 & 500 & 0.73 & 0.54 & 0.66 & 0.66 & 0.74 & 0.67 & 0.68\\
%		200 & 100 & 1.58 & 0.90 & 1.14 & 1.14 & 1.57 & 1.16 & 1.29\\
%		200 & 50  & 1.78 & 0.95 & 1.22 & 1.22 & 1.77 & 1.23 & 1.44\\
%		\hline
		Criterion       & Weighting        & \shortstack{\vspace{1mm}\\ $m=1$\\ $n=50$} & \shortstack{$m=1$\\ $n=100$} & \shortstack{$m=1$\\ $n=500$} & \shortstack{$m=200$\\ $n=100$} & \shortstack{$m=200$\\ $n=500$}\\
		\hline
		\hline
		\texttt{oracle} & \texttt{default}   & 2.39 & 2.88 & 2.89 & 1.77 & 0.72 \\
		\texttt{oracle} & \texttt{oracle}    & 2.17 & 2.49& 2.84& 1.01 & 0.55 \\
		\hline
		\texttt{oracle} & \texttt{global}    & 2.11 & 2.56 & 2.79 & 1.27 & 0.65 \\
		\texttt{oracle} & \texttt{union}     & 2.16 & 2.59 & 2.72 & 1.27 & 0.65 \\
		\texttt{oracle} & \texttt{int}       & 2.16 & 2.59 & 2.72 & 1.76 & 0.73 \\
		\texttt{oracle} & \texttt{union-int} & 2.11 & 2.58 & 2.79 & 1.28 & 0.67 \\
		\texttt{oracle} & \texttt{$\ell_1$}  & 2.23 & 2.70 & 2.82 & 1.45 & 0.67 \\
		\hline
		\hline
		\texttt{cooks} & \texttt{global}     & 2.80 & 3.01 & 3.02 & 2.15 & 0.76 \\
		\texttt{cooks} & \texttt{union}      & 2.22 & 2.69 & 2.73 & 2.15 & 0.76 \\
		\texttt{cooks} & \texttt{int}        & 2.22 & 2.69 & 2.73 & 1.76 & 0.72 \\
		\texttt{cooks} & \texttt{union-int}  & 2.85 & 2.95 & 2.98 & 1.73 & 0.72 \\
		\texttt{cooks} & \texttt{$\ell_1$}   & 2.35 & 2.82 & 2.89 & 1.74 & 0.72 \\
		\hline
		\texttt{KL} & \texttt{global}      & 2.83 & 3.00 & 3.11 & 2.03 & 0.82 \\
		\texttt{KL} & \texttt{union}       & 2.26 & 2.89 & 2.79 & 2.03 & 0.82 \\
		\texttt{KL} & \texttt{int}         & 2.26 & 2.89 & 2.79 & 1.76 & 0.74 \\
		\texttt{KL} & \texttt{union-int}   & 2.91 & 2.99 & 3.10 & 1.42 & 0.66 \\
		\texttt{KL} & \texttt{$\ell_1$}    & 2.35 & 2.82 & 2.90 & 1.54 & 0.67 \\
		\hline
		\texttt{CPO} & \texttt{global}      & 2.75 & 2.94 & 3.02 & 2.08 & 0.82 \\
		\texttt{CPO} & \texttt{union}       & 2.31 & 2.70 & 2.73 & 2.08 & 0.82 \\
		\texttt{CPO} & \texttt{int}         & 2.31 & 2.70 & 2.73 & 1.76 & 0.73 \\
		\texttt{CPO} & \texttt{union-int}   & 2.76 & 2.84 & 2.98 & 1.53 & 0.66 \\
		\texttt{CPO} & \texttt{$\ell_1$}    & 2.35 & 2.82 & 2.89 & 1.62 & 0.67 \\
		\hline
		\end{tabular}
	\end{center}
\end{table}

\begin{table}[h!]
\small
	\begin{center}
	\caption{Global RMSE performance over entire prediction domain for Friedman simulation study.}
    \label{tab:friedstudy:b}
%		\begin{tabular}{|c|c|c|c||c|c|c|c|c|c|}
%		\hline
%		$m$ & $n$ & \texttt{default} & \texttt{oracle} & \texttt{global} & \texttt{union} & \texttt{int} & \texttt{union-int} & $\ell_1$\\
%		\hline
%		\hline
%		1   & 500 & 2.77 & 2.76 & 2.77 & 2.80 & 2.80 & 2.77 & 2.77 \\
%		1   & 100 & 3.32 & 3.35 & 3.34 & 3.47 & 3.47 & 3.34 & 3.32 \\
%		1   & 50  & 3.81 & 3.81 & 3.84 & 4.05 & 4.05 & 3.84 & 3.81 \\
%		200 & 500 & 0.67 & 0.66 & 0.77 & 0.77 & 0.67 & 0.67 & 0.67 \\
%		200 & 100 & 1.50 & 1.46 & 1.64 & 1.64 & 1.50 & 1.50 & 1.50 \\
%		200 & 50  & 1.92 & 1.89 & 2.20 & 2.20 & 1.92 & 1.92 & 1.92 \\
%		\hline
		\begin{tabular}{|c|c|c|c|c|c|c|}
		\hline
		Criterion       & Weighting        & \shortstack{\vspace{1mm}\\ $m=1$\\ $n=50$} & \shortstack{$m=1$\\ $n=100$} & \shortstack{$m=1$\\ $n=500$} & \shortstack{$m=200$\\ $n=100$} & \shortstack{$m=200$\\ $n=500$}\\
		\hline
		\hline
		\texttt{oracle} & \texttt{default}   & 3.79 & 3.38 & 2.78 & 1.55 & 0.67 \\
		\texttt{oracle} & \texttt{oracle}    & 3.80 & 3.35 & 2.77 & 1.48 & 0.67 \\
		\hline
		\texttt{oracle} & \texttt{global}    & 3.81 & 3.39 & 2.78 & 1.71 & 0.78 \\
		\texttt{oracle} & \texttt{union}     & 4.03 & 3.51 & 2.81 & 1.71 & 0.78 \\
		\texttt{oracle} & \texttt{int}       & 4.03 & 3.51 & 2.81 & 1.55 & 0.67 \\
		\texttt{oracle} & \texttt{union-int} & 3.81 & 3.39 & 2.78 & 1.55 & 0.67 \\
		\texttt{oracle} & \texttt{$\ell_1$}  & 3.79 & 3.38 & 2.78 & 1.55 & 0.67 \\
		\hline
		\hline
		\texttt{cooks} & \texttt{global}     & 4.27 & 3.59 & 2.89 & 2.10 & 0.77 \\
		\texttt{cooks} & \texttt{union}      & 4.71 & 4.01 & 2.99 & 2.10 & 0.77 \\
		\texttt{cooks} & \texttt{int}        & 4.71 & 4.01 & 2.99 & 1.55 & 0.67 \\
		\texttt{cooks} & \texttt{union-int}  & 4.29 & 3.61 & 2.85 & 1.55 & 0.67 \\
		\texttt{cooks} & \texttt{$\ell_1$}   & 3.80 & 3.38 & 2.78 & 1.55 & 0.67 \\
		\hline
		\texttt{KL} & \texttt{global}      & 4.24 & 3.60 & 2.93 & 2.10 & 0.91 \\
		\texttt{KL} & \texttt{union}       & 4.67 & 4.03 & 3.05 & 2.10 & 0.91 \\
		\texttt{KL} & \texttt{int}         & 4.67 & 4.03 & 3.05 & 1.55 & 0.67 \\
		\texttt{KL} & \texttt{union-int}   & 4.25 & 3.63 & 2.88 & 1.56 & 0.67 \\
		\texttt{KL} & \texttt{$\ell_1$}    & 3.80 & 3.38 & 2.78 & 1.55 & 0.67 \\
		\hline
		\texttt{CPO} & \texttt{global}      & 4.23 & 3.61 & 2.89 & 2.09 & 0.90 \\
		\texttt{CPO} & \texttt{union}       & 4.64 & 4.04 & 2.99 & 2.09 & 0.90 \\
		\texttt{CPO} & \texttt{int}         & 4.64 & 4.04 & 2.99 & 1.55 & 0.67 \\
		\texttt{CPO} & \texttt{union-int}   & 4.21 & 3.60 & 2.85 & 1.56 & 0.67 \\
		\texttt{CPO} & \texttt{$\ell_1$}    & 3.80 & 3.38 & 2.78 & 1.55 & 0.67 \\
		\hline
		\end{tabular}
	\end{center}
\end{table}

The results are summarized in Tables \ref{tab:friedstudy:a} and \ref{tab:friedstudy:b}.  The results labeled as  \texttt{default} are regular BART without reweighting, while the \texttt{oracle} results are the best case performance achieved by explicitly training BART with the influential observation removed.  The reweighting schemes considered are labeled \texttt{global} \citep{bradlow:zaslavsky:1997}, \texttt{union} (Theorem 1), \texttt{int} (Theorem 2), \texttt{union-int} (Theorem 3) and $\ell_1$, where the $\ell_1$ method used an $L1$ distance of 0.09. % which empirically appeared to give the best performance for this approach.
A criterion setting of \texttt{oracle} denotes when the true influential observations are taken as known, whereas \texttt{cooks}, \texttt{KL} and \texttt{CPO} detects the influentials using our proposed diagnostics.

There are a few takeaways from the above study.  First, as expected, the \texttt{global} method often provides the worse performance particularly over the global prediction domain.  That is, possible improvements in local prediction near the influential observation often results in a decrease in global performance. The \texttt{union} method also displays this unfavorable tradeoff as it is most similar to the \texttt{global} method, even though the local prediction was often good. The \texttt{int} method appears to suffer from over-localization, making its performance more dependent on the behavior of the response surface and/or the settings of BART's prior. The \texttt{union-int} method appears to be the approach that is broadly robust, providing best or near-best performance in both local and global metrics.  The $\ell_1$ method can also provide good performance, but its dependence on the tuning of a distance parameter would render it computationally problematic in most cases.  Finally, while the BART \texttt{oracle} local performance remains out of reach for all methods, there is nonetheless a significant reduction in error offered by the best methods, which approach \texttt{oracle}-level performance in many cases, particularly for \texttt{CPO} with \texttt{union-int}. 

Finally, we note that the detected influentials of \texttt{cooks}, \texttt{KL} and \texttt{CPO} generally have a large degree of overlap, with perhaps some slight differences.  The most notable difference in detecting the true influentials was between \texttt{cooks} and the other methods when $m=200$ -- here, \texttt{cooks} only detected the true influentials about $5\%$ of the time while \texttt{KL} detected the influentials $70-100\%$ of the time and \texttt{CPO} achieved a perfect detection rate.  Meanwhile in the $m=1$ runs, all of the methods suffered due to the model being in the underfit regime, leading to an accuracy no higher than $45\%$ for detecting the true influentials.   This suggests combining the detected influentials amongst metrics to possibly increase performance.  We suggest combining \texttt{cooks} with \texttt{CPO} since both can choose the detection threshold in the same principled manner.

%\begin{figure}
%\centering
%\includegraphics[scale=.6]{plot_friedman_local_IS.pdf}
%\caption{Proposed resampling method ``localized'' with out-of-sample prediction points.  Note there is no difference in this case between the reweighted and unweighted predictions.}
%\end{figure}
%
%\begin{figure}
%	\centering
%	\includegraphics[scale=.6]{plot_friedman_local_IS_insample.pdf}
%	\caption{Proposed resampling method ``localized'' with in-sample prediction points.  Note the significant difference between the reweighted and unweighted predictions.}
%\end{figure}

\subsection{Real World Example}
Our motivating dataset comes from a study of biomass fuels and the application of artificial intelligence models to predicting the Higher Heating Value (HHV) of such fuels based on their molecular makeup \citep{ghugare:etal:2014}.  Biomass fuels are the fourth largest source of energy, with the most common sources being solid products such as wood and biomass pellets.  However, determining the HHV potential of a biomass fuel involves expensive and time-consuming calorimetric experiments.  Instead, a popular alternative is to use mathematical models to approximate the HHV potential of a  fuel source based on its makeup of key components.  \cite{ghugare:etal:2014} consider a dataset involving $n=536$ observations where biomass covariates recorded include the amount of carbon, hydrogen, oxygen, nitrogen and sulfur present in the fuel (as a percentage of mass), with the response being the HHV value measured in MJ/kg.  The dataset is available in the \texttt{modeldata} package on CRAN, and consists of $n=536$ samples, of which 80 are test-set observations and 456 are training-set observations.

\begin{figure}
\centering
\includegraphics[scale=0.59]{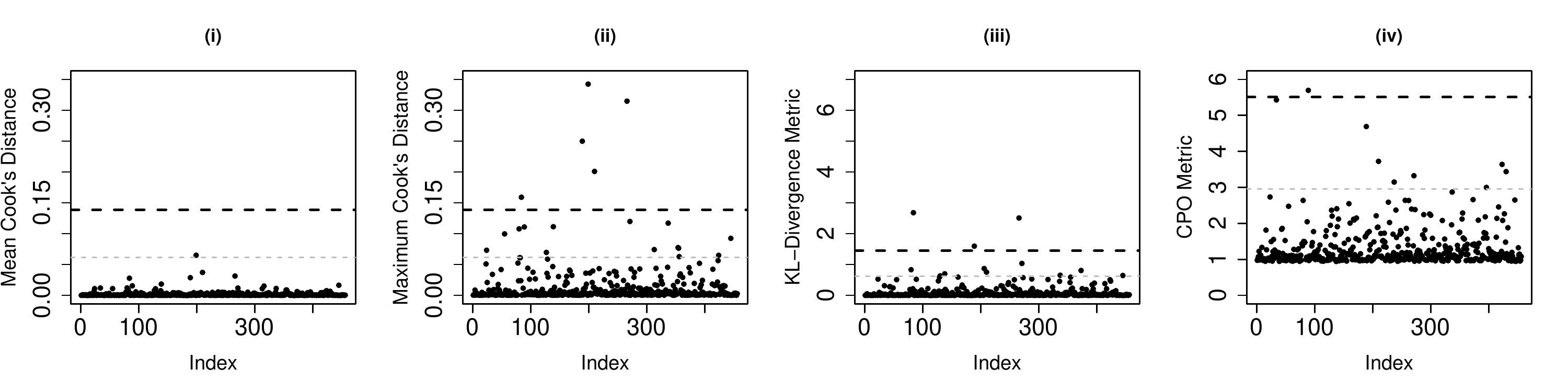}
\caption{Influence diagnostics for the HHV training data when fit using BART with $\minbot=10$ and $m=50$ trees. Panel (i) displays the mean \texttt{cooks} diagnostics, (ii) displays the maximum \texttt{cooks} diagnostic, (iii) displays the \texttt{KL} diagnostic (excluding infinities) and (iv) displays the \texttt{CPO} diagnostic (excluding infinities).  Grey dashed line denotes the $2\sigma$ cut-off while the black dashed line denotes the $3\sigma$ cut-off.}
\label{fig:biomass:influence}
\end{figure}

\begin{figure}
\centering
\includegraphics[scale=0.75]{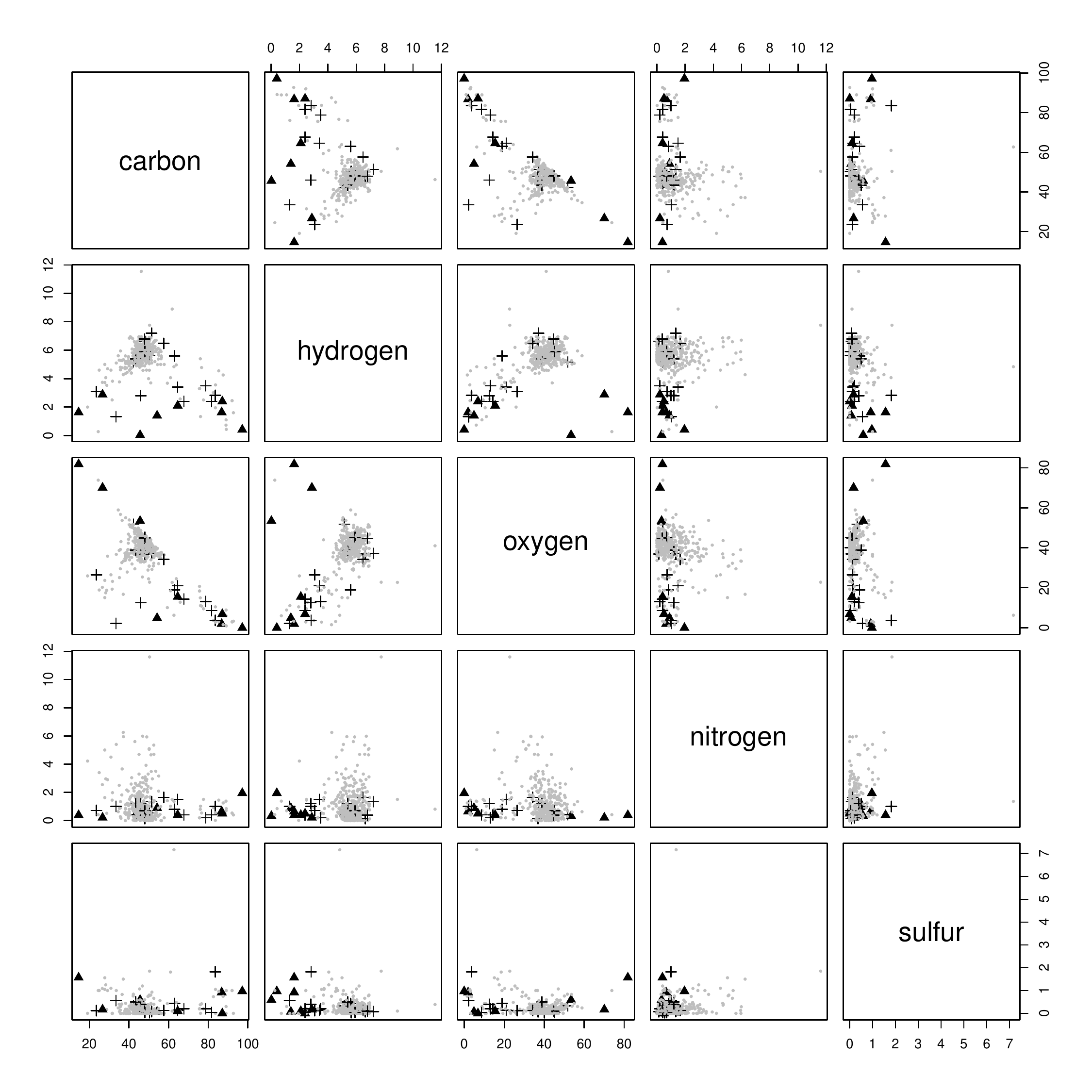}
\caption{Location of infinities (black triangles) as evaluated by the \texttt{CPO} diagnostic and additional observations marked as influentials by \texttt{KL} and \texttt{CPO} diagnostics (plus symbols) for the HHV training data (grey dots) when fit using BART.}
\label{fig:biomass:infinities}
\end{figure}

We applied BART to the training data with $\minbot=10$ and using $m=50$ trees, and explored our influence metrics to determine if there are any worriesome observations in the data.  Figure \ref{fig:biomass:influence} shows the resulting mean and maximum \texttt{cooks} diagnostics as well as the \texttt{KL} and \texttt{CPO} diagnostics.  All four metrics provide evidence of influential observations, though to varying degrees.  The mean \texttt{cooks} diagnostic seems the least sensitive in this example while the max \texttt{cooks} diagnostic is the most sensitive.  The \texttt{CPO} diagnostic is somewhere in-between these extremes, although there are additionally $8$ infinities for this metric that correspond to observations whose deletion would result in that observations terminal node failing the $\minbot$ requirement. The covariate values of observations whose \texttt{CPO} metric evaluates to infinity are shown as black triangles in Figure \ref{fig:biomass:infinities}.  As expected, these observations are located in regions of relative data sparsity and/or towards the boundaries of the range of covariate values observed.

We also note there was generally agreement about which observations were potentially problematic amongst these influence metrics. Based on this, we marked all 17 observations falling above the 2 s.d. (grey dashed) line for the \texttt{KL} and  \texttt{CPO} metric in Figure \ref{fig:biomass:influence} as influentials (note that the influentials evaluating to infinity are not shown in this panel).

\begin{table}[h!]
	\begin{center}
	\caption{RMSE performance on the HHV dataset for BART model fits as well as GP and MLP fits from \cite{ghugare:etal:2014}.}
    \label{tab:hhv}
		\begin{tabular}{|c|c|c|c|c|c|c|c|c|}
		\hline
		 & \texttt{default} & \texttt{oracle} & \texttt{global} & \texttt{union} & \texttt{int} & \texttt{union-int} & GP & MLP \\
		\hline
		\hline
		training set & 0.61 & 0.61 & 0.66 & 0.66 & 0.64 & 0.62 & 1.086 & 0.867\\
		test set & 1.49 & 1.08 & 1.60 & 1.60 & 1.32 & 1.17 & 0.942 & 0.987\\
		\hline
		\end{tabular}
	\end{center}
\end{table}

The RMSE performance of BART is summarized in Table \ref{tab:hhv}, where again \texttt{default} is the regular BART fit, \texttt{oracle} is the fit obtained by dropping the detected influentials, \texttt{global} is the reweighting method of \cite{bradlow:zaslavsky:1997}, and the remaining methods are as proposed in this paper.  In addition, the RMSE performance of \cite{ghugare:etal:2014}'s Genetic Programming (GP) and Multilayer Perceptron (MLP) models are also noted.  The performance of BART's fit on the training dataset is very strong, while the simpler reweighting methods (\texttt{global}, \texttt{union}) show a modest decrease in performance while the \texttt{int} and \texttt{union-int} methods give better results among the reweighting methods.  As in the simulation study, we again see the \texttt{union-int} demonstrating the best performance, nearly matching the in-sample performance of the regular BART fit.  In comparison, BART's performance on the test data is significantly worse than on the training data, and trails the GP and MLP models.  Again, the \texttt{union-int} method provides the highest reduction in error for BART, bringing it close to the performance of GP and MLP on the test data.  The remaining gap here could likely be explained by the smooth, continous fits of the GP and MLP models which would be a favourable characteristic for this dataset.  

Of particular interest in \cite{ghugare:etal:2014} is the performance of the models at different regimes of HHV.  In particular, they note difficulty in predicting high-HHV performance, and break down their performance summary into three ranges of HHV values: 0-16~MJ/kg, 16-25~MJ/kg and 25-36~MJ/kg.  The performance in these ranges is summarized in Table \ref{tab:hhv:rangewise}.  We see that the pattern obtained confirms \cite{ghugare:etal:2014} description of high HHV being particularly hard to predict.  Nonetheless, the \texttt{union-int} method improves on the \texttt{default} BART fit in all three regimes, and in fact beats the \texttt{oracle} performance in the 16-25~MJ/kg range where most of the observations lie.  Still, it is hard to match the performance of GP and MLP in the 0-16~MJ/kg and 16-26~MJ/kg regimes, but in the high-HHV regime the \texttt{oracle} method dominates. 

\begin{table}[h!]
	\begin{center}
	\caption{Range-wise RMSE performance on the HHV test dataset for BART model fits as well as GP and MLP fits from \cite{ghugare:etal:2014}.}
    \label{tab:hhv:rangewise}
		\begin{tabular}{|c|c|c|c|c|c|c|c|c|c|}
		\hline
		 Range & \texttt{default} & \texttt{oracle} & \texttt{global} & \texttt{union} &\texttt{int} & \texttt{union-int} & GP & MLP \\
		\hline
		\hline
		0-16MJ/kg  & 1.71 & 1.48 & 1.96 & 1.96 & 1.48 & 1.52 & 1.16 & 0.90 \\
		16-25MJ/kg & 1.02  & 1.01  & 1.15 & 1.15 & 1.03 & 1.00  & 0.84 & 0.81  \\
		25-36MJ/kg & 4.27  & 1.36 & 4.33 & 4.33 & 3.32 & 2.35  & 2.55 & 1.55  \\
		\hline
		\end{tabular}
	\end{center}
\end{table}

%% file: conclusion.tex
In this paper we proposed BART diagnostics for detecting influential observations, and devised reweighting procedures that allow posterior BART samples to be reweighted once influential observations are identified.  The influence diagnostics include a (conditional) Cook's distance metric, whose form is amenable to simple interpretation but only considers the effect of influentials on the mean function, and KL-divergence and conditional predictive distribution metrics which measure the influence of an observation on the posterior distribution.  Meanwhile, the reweighting procedures make use of importance sampling so that model training need only be done once, and the posterior samples obtained can be corrected by easily calculated weights to improve prediction performance. 

Our methods were demonstrated on both simulated data and a real-world example involving biomass fuel HHV prediction.  The consistently best method was the \texttt{CPO} diagnostic combined with the \texttt{union-int} reweighting procedure, which captures the empirical notion that highly flexible statistical learning models such as BART are affected locally by influential observations and so diagnostic and correction procedures need to capture this property in order to be practically effective.  Generally our reweighting procedure provided 10-20\% improvements in test-set prediction error as measured by RMSE while having negligible impact on training-set performance.  In contrast, directly applying global methods such as the reweighting approach of \cite{bradlow:zaslavsky:1997} significantly deteriorated both test-set and training-set performance.

Our approach has focused on prediction performance as this is perhaps the most prominent use case for BART.  Nonetheless, it would be interesting to explore extensions to alternative settings such as variable importance \citep{horiguchi:etal:2021} and high-dimensional models based on BART \citep{linero:2018}.  However, in such settings factorizing the BART posterior in a way that allows weights to be efficiently computed is likely to be problematic and a more empirical approach perhaps motivated by the $\ell_1$ method in this paper may be more practical.

Overall, we have found a suprising amount of gains can be found by addressing influential observations even though conventional wisdom suggests that highly flexible statistical learning models like BART are not affected by such problematic observations due to their localized fits.  In reality, when faced with large datasets and high-dimensional covariate spaces, the notion of `local' is very much a misnomer.  Even in 1-dimension, we can easily demonstrate the effect of influential observations on BART.  Therefore, careful application of BART should at minimum include a diagnostic step to detect possibly problematic observations, upon which investigation, removal or the reweighting procedures proposed here can be performed.

%% file: bartinfluence.bbl
\begin{thebibliography}{29}
\newcommand{\enquote}[1]{``#1''}
\expandafter\ifx\csname natexlab\endcsname\relax\def\natexlab#1{#1}\fi

\bibitem[\protect\citename{Bradlow and Zaslavsky,
  }1997]{bradlow:zaslavsky:1997}
Bradlow, E.~T. and Zaslavsky, A.~M. (1997).
\newblock \enquote{Case Influence Analysis in Bayesian Inference.}
\newblock {\em Journal of Computational and Graphical Statistics\/}, 6, 3,
  314--331.

\bibitem[\protect\citename{Breiman, }2001]{leo:2001}
Breiman, L. (2001).
\newblock \enquote{Random Forests.}
\newblock {\em Machine Learning\/}, 45, 5--32.

\bibitem[\protect\citename{Chaloner and Brant, }1988]{chaloner:brant:1988}
Chaloner, K. and Brant, R. (1988).
\newblock \enquote{A Bayesian Approach to Outlier Detection and Residual
  Analysis.}
\newblock {\em Biometrika\/}, 75, 651--659.

\bibitem[\protect\citename{Chipman et~al., }1998]{chipman:etal:1998}
Chipman, H., George, E., and McCulloch, R. (1998).
\newblock \enquote{Bayesian CART Model Search.}
\newblock {\em Journal of the American Statistical Association\/}, 93, 443,
  935--960.

\bibitem[\protect\citename{Chipman et~al., }2010]{chipman:etal:2010}
--- (2010).
\newblock \enquote{BART: Bayesian additive regression trees.}
\newblock {\em The Annals of Applied Statistics\/}, 4, 1, 266--298.

\bibitem[\protect\citename{Cook and Weisberg, }1982]{cook:weisberg:1982}
Cook, R.~D. and Weisberg, . (1982).
\newblock {\em Residuals and Influence in Regression\/}.
\newblock Chapman and Hall.

\bibitem[\protect\citename{Denison et~al., }1998]{denison:etal:1998}
Denison, D., Mallick, B., and Smith, A. (1998).
\newblock \enquote{A Bayesian CART Algorithm.}
\newblock {\em Biometrika\/}, 85, 2, 363--377.

\bibitem[\protect\citename{Gelfand et~al., }1992]{gelfand1992model}
Gelfand, A.~E., Dey, D.~K., and Chang, H. (1992).
\newblock \enquote{Model determination using predictive distributions with
  implementation via sampling-based methods.}
\newblock Tech. rep., Stanford Univ CA Dept of Statistics.

\bibitem[\protect\citename{Ghugare et~al., }2014]{ghugare:etal:2014}
Ghugare, S.~B., Tiwary, S., Elangovan, V., and Tambe, S.~S. (2014).
\newblock \enquote{Prediction of higher heating value of solid biomass fuels
  using artificial intelligence formalisms.}
\newblock {\em BioEnergy Research\/}, 7, 2, 681--692.

\bibitem[\protect\citename{Gkisser, }2017]{gkisser2017predictive}
Gkisser, S. (2017).
\newblock {\em Predictive inference: an introduction\/}.
\newblock Chapman and Hall/CRC.

\bibitem[\protect\citename{Gramacy and Apley, }2015]{gramacy:2015}
Gramacy, R.~B. and Apley, D.~W. (2015).
\newblock \enquote{Local Gaussian process approximation for large computer
  experiments.}
\newblock {\em Journal of Computational and Graphical Statistics\/}, 24, 2,
  561--578.

\bibitem[\protect\citename{Hahn et~al., }2020]{hahn2020bayesian}
Hahn, R.~P., Murray, J.~S., and Carvalho, C.~M. (2020).
\newblock \enquote{Bayesian regression tree models for causal inference:
  Regularization, confounding, and heterogeneous effects (with discussion).}
\newblock {\em Bayesian Analysis\/}, 15, 3, 965--1056.

\bibitem[\protect\citename{Hill, }2011]{hill2011bayesian}
Hill, J.~L. (2011).
\newblock \enquote{Bayesian nonparametric modeling for causal inference.}
\newblock {\em Journal of Computational and Graphical Statistics\/}, 20, 1,
  217--240.

\bibitem[\protect\citename{Horiguchi et~al., }2021]{horiguchi:etal:2021}
Horiguchi, A., Pratola, M.~T., and Santner, T.~J. (2021).
\newblock \enquote{Assessing variable activity for Bayesian regression trees.}
\newblock {\em Reliability Engineering \& System Safety\/}, 207, 107391.

\bibitem[\protect\citename{Horiguchi et~al., }2022]{horiguchi:etal:2022}
Horiguchi, A., Santner, T.~J., Sun, Y., and Pratola, M.~T. (2022).
\newblock \enquote{Using BART for Quantifying Uncertainties in Multiobjective
  Optimization of Noisy Objectives.}
\newblock {\em arXiv:2101.02558\/}.

\bibitem[\protect\citename{Johnson and Geisser, }1983]{johnson:geisser:1983}
Johnson, W. and Geisser, S. (1983).
\newblock \enquote{A predictive view of the detection and characterization of
  influential observations in regression analysis.}
\newblock {\em Journal of the American Statistical Association\/}, 78,
  137--144.

\bibitem[\protect\citename{Linero, }2018]{linero:2018}
Linero, A.~R. (2018).
\newblock \enquote{Bayesian regression trees for high-dimensional prediction
  and variable selection.}
\newblock {\em Journal of the American Statistical Association\/},  1--11.

\bibitem[\protect\citename{Liu et~al., }2020]{liu:etal:2021}
Liu, H., Nattino, G., and Pratola, M.~T. (2020).
\newblock \enquote{Sparse Additive Gaussian Process Regression.}
\newblock {\em arxiv:1908.08864\/},  1--33.

\bibitem[\protect\citename{MacKay, }1995]{mackay1995probable}
MacKay, D.~J.~C. (1995).
\newblock \enquote{Probable networks and plausible predictions-a review of
  practical Bayesian methods for supervised neural networks.}
\newblock {\em Network: computation in neural systems\/}, 6, 3, 469.

\bibitem[\protect\citename{Owen, }2013]{mcbook}
Owen, A.~B. (2013).
\newblock {\em Monte Carlo theory, methods and examples\/}.

\bibitem[\protect\citename{Pettit, }1990]{pettit1990conditional}
Pettit, L. (1990).
\newblock \enquote{The conditional predictive ordinate for the normal
  distribution.}
\newblock {\em Journal of the Royal Statistical Society: Series B
  (Methodological)\/}, 52, 1, 175--184.

\bibitem[\protect\citename{Picheny et~al., }2013]{picheny2013benchmark}
Picheny, V., Wagner, T., and Ginsbourger, D. (2013).
\newblock \enquote{A benchmark of kriging-based infill criteria for noisy
  optimization.}
\newblock {\em Structural and Multidisciplinary Optimization\/}, 48, 3,
  607--626.

\bibitem[\protect\citename{Pratola and Higdon, }2014]{Pratola:etal:2014}
Pratola, M. and Higdon, D. (2014).
\newblock \enquote{Bayesian Regression Tree Calibration of Complex
  High-Dimensional Computer Models.}
\newblock {\em Technometrics\/}.

\bibitem[\protect\citename{Pratola, }2016]{Pratola:2016}
Pratola, M.~T. (2016).
\newblock \enquote{Efficient Metropolis-Hastings Proposal Mechanisms for
  Bayesian Regression Tree Models.}
\newblock {\em Bayesian Analysis\/}, 11, 885--911.

\bibitem[\protect\citename{Starling et~al., }2020]{starling2020bart}
Starling, J.~E., Murray, J.~S., Carvalho, C.~M., Bukowski, R.~K., and Scott,
  J.~G. (2020).
\newblock \enquote{BART with targeted smoothing: An analysis of
  patient-specific stillbirth risk.}
\newblock {\em The Annals of Applied Statistics\/}, 14, 1, 28--50.

\bibitem[\protect\citename{Tan and Roy, }2019]{tan2019bayesian}
Tan, Y.~V. and Roy, J. (2019).
\newblock \enquote{Bayesian additive regression trees and the General BART
  model.}
\newblock {\em Statistics in medicine\/}, 38, 25, 5048--5069.

\bibitem[\protect\citename{Weisberg, }2013]{weisberg2013applied}
Weisberg, S. (2013).
\newblock {\em Applied linear regression\/}.
\newblock John Wiley \& Sons.

\bibitem[\protect\citename{Zellner, }1975]{zellner:1975}
Zellner, A. (1975).
\newblock \enquote{Bayesian analysis of regression error terms.}
\newblock {\em Journal of the American Statistical Association\/}, 70,
  138--144.

\bibitem[\protect\citename{Zellner and Moulton, }1985]{zellner:moulton:1985}
Zellner, A. and Moulton, B.~R. (1985).
\newblock \enquote{Bayesian regresasion diagnostics with applications to
  international consumption and income data.}
\newblock {\em Journal of Econometrics\/}, 29, 187--211.

\end{thebibliography}
